\documentclass[fleqn,usenatbib]{mnras}

\usepackage{newtxtext,newtxmath}
\usepackage{xcolor}

\usepackage[T1]{fontenc}

\DeclareRobustCommand{\VAN}[3]{#2}
\let\VANthebibliography\thebibliography
\def\thebibliography{\DeclareRobustCommand{\VAN}[3]{##3}\VANthebibliography}

\usepackage{graphicx}	
\usepackage{amsmath}	

\usepackage{amsmath}

\graphicspath{{./}{figures/}}

\newcommand{\SkyLine}{\textsc{SkyLine}\,}
\newcommand{\synsky}{\textsc{Agora}\,}
\newcommand{\universemachine}{\textsc{UniverseMachine\,}}
\title[Multi-tracer LIM lightcones]{A  multi-tracer empirically-driven approach to line-intensity mapping lightcones}
\author[Sato-Polito, Kokron, Bernal]{Gabriela Sato-Polito$^{1}$\thanks{\href{mailto:gsatopo1@jhu.edu}{gsatopo1@jhu.edu}}, 
Nickolas Kokron$^{2}$\thanks{\href{mailto:kokron@stanford.edu}{kokron@stanford.edu}},
José Luis Bernal$^{3,1}$\thanks{\href{mailto:jlbernal@mpa-garching.mpg.de}{jlbernal@mpa-garching.mpg.de}}\\
$^{1}$William H. Miller III Department of Physics and Astronomy, The Johns Hopkins University, Baltimore, MD 21218,\\
$^{2}$Kavli Institute for Particle Astrophysics and Cosmology and Department of Physics, Stanford University, Stanford, CA, USA,\\
$^{3}$Max-Planck-Institut f\"ur Astrophysik, Karl-Schwarzschild-Str. 1, 85748 Garching, Germany
}
\begin{document}
\maketitle
\begin{abstract}
Line-intensity mapping (LIM) is an emerging technique to probe the large-scale structure of the Universe. By targeting the integrated intensity of specific spectral lines, it captures the emission from all sources and is sensitive to the astrophysical processes that drive galaxy evolution. Relating these processes to the underlying distribution of matter introduces observational and theoretical challenges, such as observational contamination and highly non-Gaussian fields, which motivate the use of simulations to better characterize the signal. In this work we present \SkyLine, a computational framework to generate realistic mock LIM observations that include observational features and foreground contamination, as well as a variety of self-consistent tracer catalogs. We apply our framework to generate realizations of LIM maps from the \textsc{MultiDark Planck 2} simulations coupled to the \textsc{UniverseMachine} galaxy formation model. We showcase the potential of our scheme by exploring the voxel intensity distribution and the power spectrum of emission lines such as  21 cm, CO, {[CII]}, and Lyman-$\alpha$, their mutual cross-correlations, and cross-correlations with galaxy clustering. We additionally present cross-correlations between LIM and sub-millimeter extragalactic tracers of large-scale structure such as the cosmic infrared background and the thermal Sunyaev-Zel'dovich effect, as well as quantify the impact of galactic foregrounds, line interlopers and instrument noise on LIM observations. These simulated products will be crucial in quantifying the true information content of LIM surveys and their cross-correlations in the coming decade, and to develop strategies to overcome the impact of contaminants and maximize the scientific return from LIM experiments. 

\end{abstract}

\section{Introduction} \label{sec:intro}
Line-intensity mapping (LIM) measures the integrated emission from all galaxies and diffuse intergalactic medium (IGM) along the line of sight, surveying large cosmological volumes with line-of-sight tomographic information from targeting known spectral lines at different frequencies~\citep{Kovetz:2017agg, Liu:2019awk, Bernal:2022jap}. Eschewing expensive, high-significance resolved detections, enables the use of low-aperture instruments with modest experimental budgets, and extends the reach of the survey to higher redshifts than wide galaxy surveys. Large-scale intensity fluctuations follow matter overdensities, while the amplitude of the intensity field and the line ratios depend on the astrophysical properties of galaxies and the IGM~\citep{2019ARA&A..57..511K}. 

Numerous LIM experiments are currently underway~\citep{DeBoer:2016tnn, MeerKLASS:2017vgf, Keating:2016pka,Keating:2020wlx, Cleary:2021dsp, 2020A&A...642A..60C, Gebhardt:2021vfo} or are expected to see first light in the next few years~\citep{CCAT-Prime:2021lly, Sun:2020mco, 2021JATIS...7d4004S, 2020arXiv200914340V, 2014arXiv1412.4872D}, targeting a variety of spectral lines spanning from radio to near ultraviolet (UV) wavelengths sourced at redshifts that range from today to cosmic dawn. As the experimental prospects broaden, a variety of questions have emerged regarding, e.g., the sensitivity of LIM technique, optimal summary statistics, line-intensity modeling, and strategies to minimize the impact of observational contaminants. 

Astrophysical dependence introduces additional non-Gaussianity to the observed line-intensity maps with respect to other probes of large-scale structure, such as galaxy number counts. Along with nonlinear clustering and non-trivial observational effects, this motivates the use of LIM mocks to simulate observations that address the questions listed above.  

There are different approaches to simulate intensity maps, with a wide spectrum in the compromise between volume, accuracy in the clustering and astrophysics, and computational requirements. Most often, LIM mocks are obtained by assigning line luminosities to halo catalogs (i.e., painting), which may come from N-body simulations or approximate methods. On one end, lognormal simulations paired with input luminosity functions to assign line-luminosities to each particle maximize the performance at the cost of accuracy in the astrophysical modeling and clustering~\citep{Ramirez-Perez:2021cpq}. Otherwise, halo-finding (and associated measurements of masses) is usually required. With the exception of scaling relationships directly connecting line luminosities to halo masses (see e.g.,~\citet{Villaescusa-Navarro:2018vsg, COMAP:2021rny, 2022ApJ...929..140Y}), astrophysical properties such as the star-formation rate must be assigned to all halos first. This is commonly done by parameterizing mean relationships  between halo mass and star formation rate, as well as including a scatter to account for population variability~\citep{Silva:2014ira,Yue:2015sua,Li:2015gqa,COMAP:2018svn,COMAP:2018kem,2020MNRAS.496L..54M,MoradinezhadDizgah:2021dei,Murmu:2021quo, COMAP:2021rny}. Recent advances in studies of the galaxy--halo connection~\citep{Wechsler:2018pic} have led to the popularization of models which account for the relationship between dark matter halos and the astrophysical properties of their constituents in more robust ways. For example, augmenting halo catalogs with empirical techniques such as abundance matching ~\citep{Conroy_2006,Bethermin:2017ngy, 2019MNRAS.488.3143B, Trac:2021qbn, Bethermin:2022lmd, Gkogkou:2022bzo}, that can describe a large range of observations across cosmic time after being calibrated to a subset of observables like the luminosity function. Once halos have been assigned astrophysical properties, additional scaling relations (calibrated from observations or ab-initio simulations) may be used to paint any signal of interest onto the catalogs. 

The astrophysical dependence of the LIM signal can be modeled with more detail, yielding self-consistent predictions for several spectral lines. One approach evolves the cosmological growth of structures and the ISM and IGM astrophysics at the same time using perturbation theory, excursion-set formalism and galaxy-formation models~\citep{Mesinger:2007pd, Mesinger:2010ne, Parsons:2021qyw, Mas-Ribas:2022jok, Sun:2022ucx}. On the other hand,  semi-analytic models~\citep{2015ARA&A..53...51S} can be adapted to simulate multiple line luminosities for each galaxy~\citep{Lagos:2012sv, 2016MNRAS.461...93P, Dumitru:2018tgh, 2018A&A...609A.130L,2019MNRAS.482.4906P,  2020ApJ...905..102L}. Finally, full-fledged  hydrodynamic simulations~\citep{Hopkins:2013vha,McAlpine:2015tma,2015ApJ...813...36V,2017ApJ...846..105O,Nelson:2018uso,Villaescusa-Navarro:2018vsg, Dave:2019yyq,Pallottini:2022inw, Kannan:2021xoz} can be post-processed to consistently predict the luminosity of several lines~\citep{2020MNRAS.492.2818L, 2021arXiv211105354S, Kannan:2021ucy}, including radiative transfer, although for very limited volumes. 

Most LIM mocks and simulations are limited to small volumes or areas on the sky, with the exception of analyses that paint the signal to snapshots of simulations as opposed to lightcones. Nonetheless, lightcones are key for LIM mocks: LIM experiments often probe broad redshift ranges, and therefore the redshift evolution of clustering and astrophysical properties over the observed volume must be taken into account. Furthermore, using a lightcone is the only way to properly account for line-interlopers (different spectral lines that are redshifted into the same observed frequency band) and the effect of continuum foregrounds along the line of sight. Finally, narrow-field observations (and, similarly, small-volume simulations) are limited by cosmic variance uncertainties. The effects of cosmic variance are more severe for luminosity-dependent observables than for e.g., the galaxy power spectrum, as explored analytically in \citet{Sato-Polito:2022fkd} for the distribution of measured intensities within voxels, and in \citet{2020ApJ...904..127K} and \citet{Gkogkou:2022bzo} for the luminosity function and LIM power spectrum (only in the latter).  Proper accounting for cosmic variance and overcoming the associated limitations is an additional motivation for large-volume LIM mocks.

\begin{figure*}
    \centering
    \includegraphics[width=\textwidth]{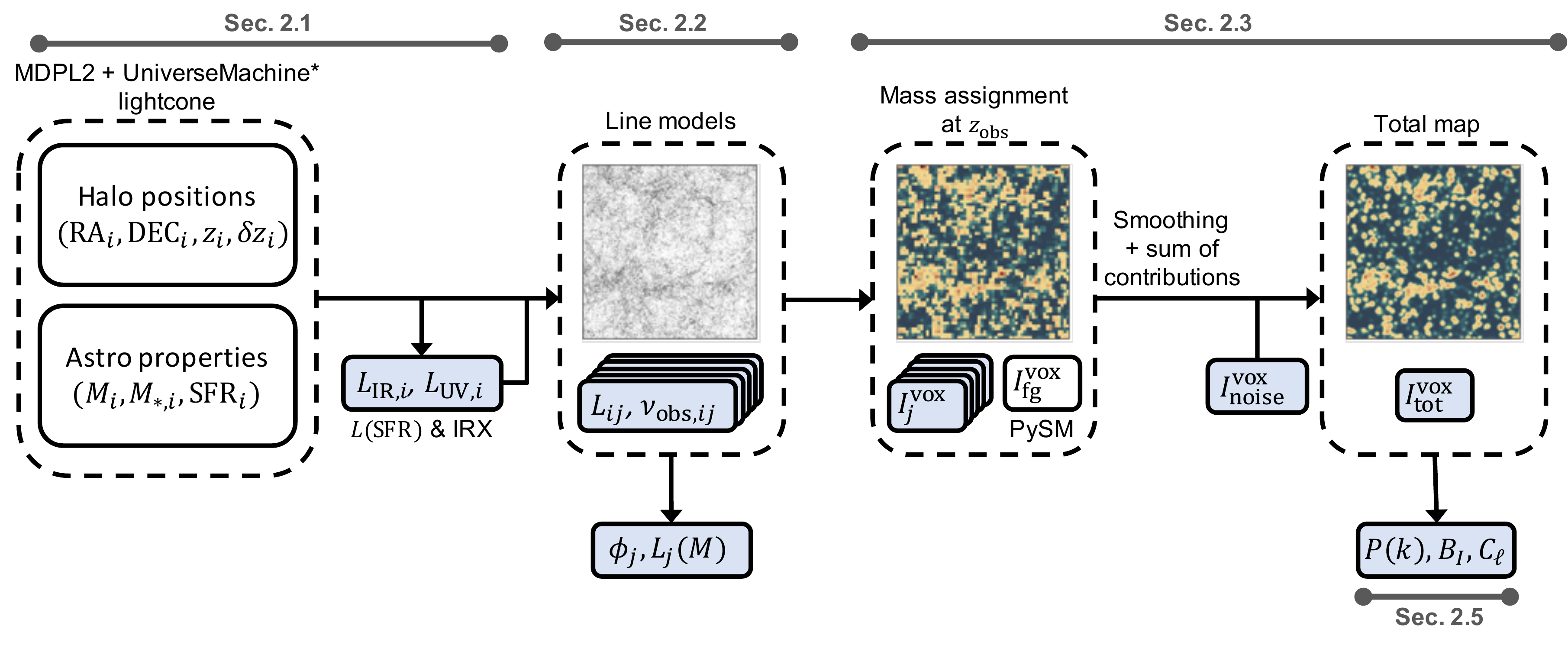}
    \caption{Summary of the steps (ordered from left to right) taken by \SkyLine to create mock line-intensity maps. Blue boxes mark quantities computed within \SkyLine, arrows denote quantities needed for the next step and the lowest row contains summary statistics computed from the quantities just above them. Each halo and spectral line are labeled with $i$ and $j$, respectively. The mass assignment at $z_{\rm obs}$ accounts for line interlopers interpreted as if they had the rest-frame frequency of the target line, and can result in three-dimensional (with voxels) or angular (with pixels) maps. The default is to use halo positions and astrophysical properties from \textsc{MDPL2} and \textsc{UniverseMachine}, although the astrophysical properties can also be read from interpolating tables or fitting functions.}
    \label{fig:diagram}
\end{figure*}
In this work we present \SkyLine\footnote{{\url{https://github.com/kokron/skyLine}}}, a numerical scheme to generate LIM mocks of any spectral line sourced within collapsed dark matter halos (thereby excluding pre-reionization neutral hydrogen in the intergalactic medium). Our starting point are the publicly available MultiDark Planck 2 (MDPL2)~\citep{Klypin:2014kpa, Rodriguez-Puebla:2016ofw} simulations and galaxy catalogs derived from the \textsc{UniverseMachine}~\citep{2019MNRAS.488.3143B} galaxy formation model, which we use to generate a lightcone. We adopt an empirical approach akin to \citet{Li:2015gqa} and \citet{Chung:2018szp} in order to assign line luminosities from halo or galaxy properties. Semi-analytic models and hydrodynamic simulations are complementary to empirical techniques; the former rely on physical assumptions regarding the formation and evolution of galaxies and gas, which may lead to more rigid predictions of observational signals. Meanwhile,  empirical techniques offer greater flexibility, which may be particularly revealing when applied to highly uncertain astrophysical processes, as is the current state-of-affairs of intensity mapping. 

Beyond the signal from the targeted spectral lines, a variety of other sources of intensity are included. The contributions from line-interlopers and foregrounds are modeled and projected into the maps, which can be smoothed with given angular and spectral resolutions. The instrument noise is then added to produce the final three-dimensional or angular observed intensity map. In addition, maps of number counts of galaxies, distinguishing between quenched and star-forming populations, can be generated.  Ultimately, the tool presented in this work offers a playground to realize realistic mock observations of line intensities, complementing ongoing efforts to jointly model and interpret multi-line observations ~\citep{2019ApJ...887..142S, Mas-Ribas:2022jok, Sun:2022ucx}.

Ongoing and planned LIM experiments will not only overlap with each other, but also with galaxy surveys and CMB observations (see~\citet{Bernal:2022jap}). Cross-correlations between LIM and other cosmological probes have been considered both as a tool to disentangle the signal from a noisy data set, leading to the first reported detections~\citep{Chang:2010jp, BOSS:2015ids, CHIME:2022kvg, Cunnington:2022uzo}, and as complementary probe of large-scale structure. To explore this possibility, we have constructed fiducial catalogs from the same large-scale structure and galaxy formation model as the \synsky~\citep{Omori:SimSky}. This mutual compatibility enables, for the first time, a coherent multi-probe study of the CMB, its secondaries and large-scale structure observables with self-consistent and accurate clustering and star formation. 

We present the methodology to produce the final line-intensity or galaxy number count maps and the summary statistics considered in \S~\ref{sec:meth}; the results and analysis of line-intensity maps in \S~\ref{sec:analysis}; the cross-correlations with other observables in \S~\ref{sec:cross}; and conclude in \S~\ref{sec:conclusion}. Note that throughout this work, we discuss the default settings of \SkyLine, but the code is highly customizable and its modular structure is designed to encourage extensions.
\section{Methodology}
\label{sec:meth}
In this section we describe the methodology employed to generate the mock line-intensity and galaxy distribution maps. A summary diagram can be found in Fig.~\ref{fig:diagram}.

\subsection{Halo catalogue and astrophysical properties}
\label{sec:halo_astro}
We begin by selecting an input halo catalog. The minimum information required to produce a simulated LIM map are the angular positions of halos on the sky (right ascension RA and declination DEC), their redshift $z$, and halo mass $M$, along with a relation that connects halo mass with line luminosity. The connection between halo and spectral line may, for example, rely on external relationships to other relevant astrophysical properties such as star-formation rate SFR, stellar mass $M_*$, and infrared (IR) luminosity ($L_{\rm IR}$). 

The fiducial simulations we present here are derived from the publicly available \textsc{MDPL2}+\textsc{UniverseMachine} catalogs\footnote{\href{https://www.peterbehroozi.com/data.html}{https://www.peterbehroozi.com/data.html}}. \textsc{MDPL2} is a $V=$1~$h^{-3}$Gpc$^3$ dark-matter-only N-body simulation that evolves 3840$^3$ particles, corresponding to a mass resolution of $1.5\times 10^9$ $h^{-1}M_{\odot}$. Its force resolution is 5~$h^{-1}$kpc, and assumes cosmology consistent with the results of \citet{Planck:2015fie}. The halo finding and merger tree construction were performed using \textsc{Rockstar}~\citep{2013ApJ...762..109B} and \textsc{Consistent Trees}~\citep{2013ApJ...763...18B} algorithms, respectively. \textsc{UniverseMachine} then produces an empirically-driven star-formation history for each dark matter halo. 

The publicly available \textsc{MDPL2-UM} data products consist of halos distributed across 126 snapshots, from redshifts between $z=0-15$; however, we only produce lightcones out to $z=9.7$ in this work. At higher redshifts the limited mass resolution of \textsc{MDPL2-UM} implies a significant portion of the total SFR is sourced by halos below the resolution limit. We investigate the impact of resolution as a function of redshift in Appendix~\ref{appendix:A}, but note that at $z\sim5$ our fiducial simulation captures a large fraction of the total star formation. Additionally, the emission of many lines is suppressed for small halos, and the total line luminosity will be less suppressed than the corresponding amount of star-formation in the Universe. Finally, we note that there is a dearth of observations to constrain \universemachine at higher redshifts, but we expect this to change rapidly in the era of JWST.

We proceed to create the raw, unprocessed, lightcone halo catalogs following the same procedure as \citet{Omori:SimSky}. That is, we define an observer at the center of a tiling of the $z\approx0$ snapshots. Spherical shells of width $\Delta \chi = 25 h^{-1} {\rm Mpc}$ are grown out from the observer. The snapshots with a scale factor closest to the center of the shell are loaded, and halos that intersect the shell are added to the lightcone catalogs. In order to prevent the repetition of structures as the radius (and volume) of these shells increase, the tiling of snapshots is rotated every $\chi = 1000 h^{-1} {\rm Mpc}$. This rotation erases spatial correlations beyond $1000 h^{-1} {\rm Mpc}$, but we will focus on significantly smaller scales, which are unaffected by the rotation. These are the same rotations as those used in \cite{Omori:SimSky}, as well as the same underlying \textsc{MDPL2-UM} snapshots. This choice ensures that our final line-intensity maps will be self-consistently correlated by design with the products of~\citet{Omori:SimSky}. Given a catalog of halos, subhalos and their associated \textsc{UniverseMachine} star-formation rates and stellar masses, our scheme is now capable of deriving line luminosities for every entry in the catalog. We additionally produce angular positions, redshifts, and distortions $\Delta z$ due to radial peculiar velocities.

For some line luminosities, it is also desirable to derive an IR luminosity from the underlying SFR and stellar masses. We do this by considering the balance between UV and IR radiation: radiation in the UV range is produced primarily by massive short-lived stars, the population of which can be approximated with the SFR, and is partially absorbed and re-emitted in the IR by dust. The total SFR can then be understood as the sum of the contributions that can be inferred from the observed rest-frame UV and the dust-obscured component inferred from IR, such that
\begin{equation}
    \text{SFR} = K_{\rm UV} L_{\rm UV} + K_{\rm IR} L_{\rm IR}\,,
    \label{eq:SFR_KS}
\end{equation}
where $L_{\rm UV}$ is the rest-frame escaped UV luminosity. We use $K_{\rm UV} = 2.5\times 10^{-10}\ M_{\odot} {\rm yr}^{-1} L^{-1}_{\odot}$ and $K_{\rm IR} = 1.73\times 10^{-10}\ M_{\odot} {\rm yr}^{-1} L^{-1}_{\odot}$ from \citet{Madau:2014bja}, after multiplying them by a 0.63 factor to convert them to the Chabrier initial mass function \citep{2003PASP..115..763C}, consistent with \textsc{UniverseMachine}. The ratio between IR and UV luminosities observed in galaxies is known as the infrared excess (IRX$\equiv L_{\rm IR}/L_{\rm UV}$); we use the the parameterization as a function of stellar mass $M_{*}$ presented by \citet{2020ApJ...902..112B}
\begin{equation}
    \text{IRX} = \left(\frac{M_*}{M_s}\right)^{\alpha_{\rm IRX}},
    \label{eq:IRX}
\end{equation}
with best-fit values of $\log_{10}(M_s/M_{\odot}) = 9.15^{+0.18}_{-0.16}$ and $\alpha_{\rm IRX}=0.97^{+0.17}_{-0.17}$, with a log-normal scatter of $\sigma_{\rm IRX} = 0.2$.\footnote{The IRX from~\citet{2020ApJ...902..112B} is calibrated at 1600 \AA, while $K_{\rm UV}$ in Eq.~\eqref{eq:SFR_KS} assumes UV radiation at 1500 \AA. For a typical spectral slope $\beta=-2$ of the UV continuum (see Fig.~2 in~\citet{2020ApJ...902..112B}) this corresponds to a $\sim 14\%$ change in $L_{\rm UV}$. We do not account for this change, since it is effectively absorbed by theoretical and observational uncertainties in the constant $K_{\rm UV}$.} Alternative IRX relations with the stellar mass~\cite{Heinis:2013dsa, 2016ApJ...833...72B} are also implemented. Following the approach outlined in \citet{Wu:2016vpb}, our final expression connecting SFR to $L_{\rm IR}$ is
\begin{equation}
    L_{\rm IR} = \frac{\rm SFR}{K_{\rm IR} + K_{\rm UV}{\rm IRX}^{-1}}\,,
    \label{eq:LIR}
\end{equation}
which results in a lower $L_{\rm IR}$ for light stellar mass halos with respect to the simpler Kennicutt-Schmidt relation \citep{1998ApJ...498..541K}. Similarly, we can use Eqs.~\eqref{eq:IRX} and~\eqref{eq:LIR} to obtain $L_{\rm UV}$ for each halo. \par 
{The fiducial ${\rm IRX} (M_*)$ relation adopted was chosen to coincide with the implementation of the \synsky simulations. \citet{Omori:SimSky} verified 
the observables 
which depend on IR luminosity 
against observations of cosmic star formation density and cosmic infrared background $\times$ CMB cross-correlations and properties of IR point sources. In the original work which presents this infrared excess \citep{2020ApJ...902..112B} additional comparisons are made to collections of IR luminosity data (see their figures 15 and 19). Nevertheless, the other parameterizations between IR luminosity and star-formation rate may be implemented. In the public release of \SkyLine we also include the original formulation from \cite{Li:2015gqa} and the most recent parameterization used in analysis of data from the COMAP collaboration which depends solely on host halo mass~\citep{COMAP:2021rny}.}

{The starting point of our mock maps relies on a semi-empirical model, which may be insufficient to model certain LIM signals due to the limited number of galaxy properties captured. In particular, the metallicity and gas fraction can affect usual target spectral lines, especially at high redshifts and low-mass systems, which usually lie outside of the applicability of many scaling relationships due the scatter and lack of observations. An example includes the so-called [CII] deficit (see e.g.,~\cite{Liang:2023sxx}). Future developments of \SkyLine will improve the line-emission modeling in this direction, following different complementary approaches, such as theoretically motivated analytic emission models (see e.g.,~\cite{2019ApJ...887..142S}), scaling relationships relating stellar mass and metallicity (see e.g.,~\citet{Maiolino:2008gh}),  conditional probability distributions of galaxy properties~\citep{Zhang:2023oem}, line broadening~\citet{COMAP:2021rny},  informed models of line-emission from more focused, smaller volume, simulations (see e.g.,~\citet{Kannan:2021ucy}), etc.}

\subsection{Line-luminosity modeling}
With the exception of HI emission before reionization, i.e., at $z\gtrsim 5-6$, (and, to a lesser degree, part of the Lyman-$\alpha$ emission, see e.g.,~\citet{Niemeyer:2022vrt, Niemeyer:2022arn}), all spectral lines originate in galaxies and the IGM within dark matter halos. Empirical relations connecting halos and line luminosity can therefore be used to predict nearly all lines of interest and present a unique opportunity for multi-tracer LIM analyses. For each line,  we include a wide variety of scaling relationships as a function of the astrophysical quantities described in the previous section, with free parameters that can be easily adapted to different calibrations, or extended if needed. We briefly describe below the main set of relationships included. Other theory-motivated models, aiming to self-consistently model several spectral lines from a simple model for the IGM and ISM (see e.g.,~\citet{2019ApJ...887..142S}) can easily be implemented. 

The linear Kennicutt-Schmidt  relation between line luminosity and SFR usually applies to accurate tracers of star formation, such as the Balmer H$\alpha$ and H$\beta$ lines and the optical {[OII]} ({3727}\AA) and {[OIII]} ({5007}\AA) lines. For those cases, and including magnitude-averaged mean dust-extinction laws, we have~\citep{Kennicutt:1998zb,Villa-Velez:2021ojy}
\begin{equation}
    \frac{L}{L_{\odot}} = \alpha_K\times\frac{{\rm SFR}}{M_\odot{\rm yr}^{-1}}\times10^{-A_{\rm ext}/2.5} \,,
\end{equation}
where $A_{\rm ext}$ is the extinction parameter for each line. The Lyman-$\alpha$ (Ly$\alpha$) line luminosity follows a similar relation, with a more complicated dependence on the escape fraction $f_{\rm esc}\equiv 10^{-A_{\rm ext}/2.5}$ due to more complex radiative transfer and more severe extinction.

Additional parameterizations often involve power-laws, either with the SFR,
\begin{equation}
    \log_{10}\frac{L}{L_\odot} = \alpha_{\rm SFR}\log_{10}\frac{{\rm SFR}}{M_\odot{\rm yr}^{-1}} + \beta_{\rm SFR}\,,
    \label{eq:SFRtoL_PL}
\end{equation}
or the infrared luminosity,
\begin{equation}
    \log_{10}\frac{L}{L_\odot} = \alpha_{\rm IR}\log_{10}\frac{L_{\rm IR}}{L_\odot} + \beta_{\rm IR}\,.
    \label{eq:LtoLIR}
\end{equation}
While the emission of {[CII]} presents correlations with both the SFR and the far IR dust emission, other fine-structure dust-cooling emission lines and the CO rotational lines correlate mostly with the latter~\citep{Spinoglio:2011ug}. 
Another possibility is to directly model the relation between line-luminosity and halo mass 
\begin{equation}
     \frac{L}{L_{\odot}} = \gamma_M f(M),
\end{equation}
where $\gamma_M$ is an dimensionful constant that depends on the specific line and form of $f(M)$. Examples of the function $f$ that have been previously adopted in the literature are single and double power laws, as well as the inclusion of an exponential suppression for low masses. For the HI line at 21 cm after reionization, for example, the line luminosity is directly related to the neutral hydrogen mass, which can be parametrized as~\citep{Villaescusa-Navarro:2018vsg}
\begin{equation}
    M_{\rm HI}(M) = M_0 \left(\frac{M}{M_{\rm min}}\right)^{\alpha_{\rm M}} \exp\left\{ -\left(\frac{M_{\rm min}}{M}\right)^{\beta_{\rm M}}\right\},
    \label{eq:LofM_HI}
\end{equation}
and in this case coefficient $\gamma_M$ converts the neutral hydrogen mass enclosed in the halo to its luminosity (see Eqn.~\ref{eqn:lum_HI}).

Beyond atomic and molecular lines, exotic emission lines coming from e.g., dark matter decays \citep{Bernal:2020lkd} or decays from the cosmic neutrino background \citep{Bernal:2021ylz} can be also implemented within the framework presented here.

\subsection{Intensity map-making}
\begin{figure*}
    \centering
    \includegraphics[width=\textwidth]{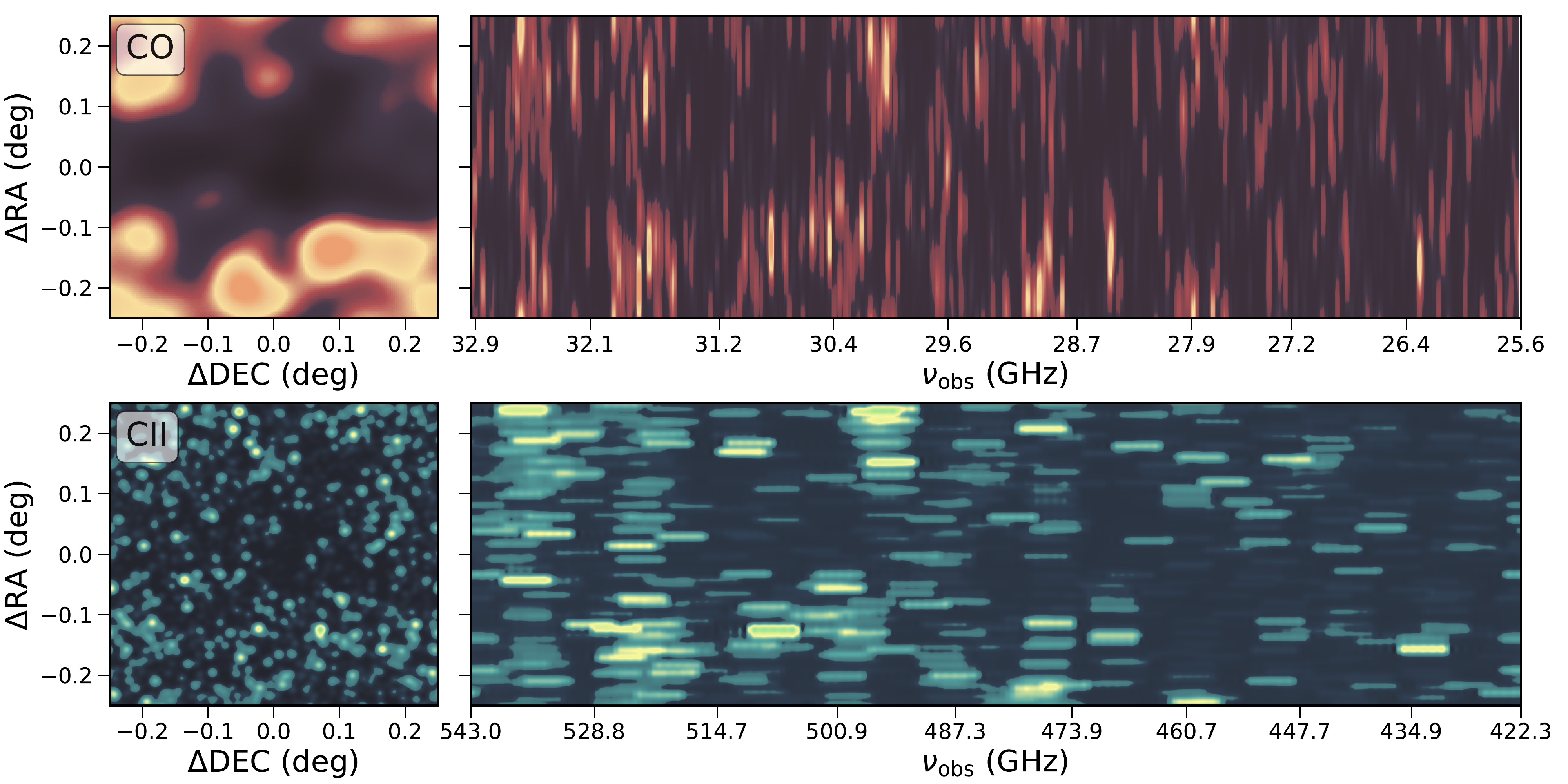}
    \caption{Three-dimensional noise-less mock CO (top) and {[CII]} (bottom) intensity maps, for observed frequencies such as they correspond to $z\in \left[2.5,3.5\right]$ in each case. The panels on the left are projected along the line of sight and the ones on the right are projected along the x-axis. We apply COMAP-like ($\theta_{\rm FWHM}=4'$, $\delta\nu=31.25\,{\rm MHz}$) and FYST-like ($\theta_{\rm FWHM}=35''$, $\delta\nu=5.5\,{\rm GHz}$) resolutions. Both maps have their mean removed, and brighter colors correspond to the higher intensity in a linear colorbar.}
    \label{fig:example_map}
\end{figure*}

In order to generate line-intensity maps we must compute the intensity in each angular and frequency bin of the considered experiment. We consider contributions to the observed intensity from the target and interloper lines, continuum foregrounds, and instrumental noise. 

Throughout this section we use the  experimental angular and spectral resolutions, given by the  full-width half maximum $\theta_{\rm FWHM}$ telescope beam and the frequency-channel width $\delta\nu$ (or the minimum frequency used for science studies) to set the map resolution along the angular and radial directions, respectively. From that base, we will consider supersample factors $N_{\rm ss,\theta}$ and $N_{\rm ss, \nu}$ (integers larger or equal to unity) in case we require higher resolution in angle or frequency. 

\subsubsection{Spectral lines}
\label{sec:map_lines}
First, we filter the halo catalog, leaving only the halos that fall within the observed footprint and frequency bandwidth of the experiment in consideration. The last condition depends not only on the redshift but also on the lines considered, since for the $i$-th halo the observed frequency $\nu_{ij,{\rm obs}}$ of the $j$ line with rest-frame frequency $\nu_j$ is
\begin{equation}
    \nu_{ij,{\rm obs}} = \frac{\nu_j}{1+z_i+\Delta z_i}\,,
\end{equation}
where $z_i$ and $\Delta z_i$ are the cosmological redshift of the $i$-th halo and its distortion due to peculiar velocities along the line of sight, respectively. That is, the peculiar motion of the halo can shift a line outside of the observed frequency range. 

We assume that all line profiles are delta functions centered on $\nu_j$,  neglecting line-broadening due to the computational challenge of directly including this effect in our simulation. An approximate modelling would entail a Gaussian smoothing of the emission along the line-of-sight (frequency) direction for each individual halo using the rotation velocities obtained from the N-body simulation (see~\citet{COMAP:2021rny} for more details).

Then we populate each voxel with emitters, a procedure which depends on whether we are generating an angular or a three-dimensional map. For the angular map, we use a \textsc{HEALpix} tessellation of the celestial sphere \citep{Gorski_2005}, using the python wrapper \textsc{healpy} \citep{2020ascl.soft08022Z}, with pixel area smaller than $\left(\theta_{\rm FWHM}/N_{\rm ss,\theta}\right)^2$, and mask all pixels outside of the area probed by the simulated experiment.

The production of three-dimensional maps requires the conversion of observed frequencies and angles into physical distances. In order to fully account for the projection effects for line interlopers~\citep{Gong:2013xda, Lidz:2016lub, 2017ApJ...835..273G, Cheng:2016yvu, Gong:2020lim}, we choose a target spectral line and use its rest-frame frequency to transform \textit{all} observed frequencies to redshifts. Then, assuming the $\Lambda$CDM best-fit cosmology of~\cite{Planck:2015fie} we transform angular coordinates and redshifts into spatial coordinates. The physical position of each emitter and its contribution to the signal are added to a grid using \textsc{nbodykit} \citep{Hand:2017pqn} using cloud-in-cell interpolation, with a separate map per line. As most LIM surveys will survey relatively small footprints on the sky, the conversion of a data-cube to physical coordinates introduces minimal distortions (such as wide-angle effects); further optimization and effects from the mask can be characterized using suitable weights with a FKP-like estimator \citep{Feldman_1994,Yamamoto_2006,Blake_2019}. For the rest of the publication, however, we will mitigate the impacts of the lightcone geometry on our 3D intensity maps by extracting the largest rectangular cuboid inscribed within the lightcone. 

The aggregated intensity within each voxel for the line $j$ is then
\begin{equation}
\begin{split}
    I^{\rm vox}_j = & \sum_{i_{{\rm vox},j}}\frac{c}{4\pi \nu_j H(z_{i_{{\rm vox},j}})}\rho_{L, j} \\ = & \sum_{i_{{\rm vox},j}}\frac{c}{4\pi \nu_j H(z_{i_{{\rm vox},j}})}\frac{L_{{i_{{\rm vox},j}}j}}{V_{{\rm vox},j}},
\label{eq:LtoI}
\end{split}
\end{equation}
where ${i_{{\rm vox},j}}$ indexes the halos that are in the voxel of interest for the line $j$, $c$ is the speed of light, $H$ is the Hubble parameter, and $\rho_{L, j}$ is the the local luminosity density of the $j$ line, which we define as the sum of the luminosities produced by each halo divided by the voxel volume $V_{{\rm vox},j}$ (which depends on the resolution and redshift considered). For angular maps, we consider that the radial size of the voxel is given by the frequency-channel width and integrate the signal across all frequency bins. 

For experiments observing below some tens of GHz, the signal is often interpreted as brightness temperature $T$ by using the Rayleigh-Jeans relation
\begin{equation}
    T_j = \frac{c^2 I_{j}}{2 k_{\rm B} \nu^2_{j,{\rm obs}}},
    \label{eq:RJ}
\end{equation}
where $k_{\rm B}$ is the Boltzmann constant. The total signal for each angular and frequency bin is the sum of all contributions from each line. 

Finally, the limited angular and spectral resolutions of LIM experiments result in a smearing of small-scale fluctuations on directions perpendicular and parallel to the line-of-sight, respectively. To account for this effect, we consider the following characteristic scales
\begin{equation}
    \sigma_{\perp} = \chi(z_{{\rm true},j}) \sigma_{\rm beam}, \qquad    \sigma_{\parallel} = \frac{c(1+z_{{\rm true},j}) \delta{\nu}}{H(z_{{\rm true},j}) \nu_{\rm obs}}
\end{equation}
transverse and along the line of sight, respectively, where $\sigma_{\rm beam} = \theta_{\rm FWHM}/\sqrt{8 \log 2}$, $\chi$ is the comoving radial distance and $z_{{\rm true},j}$ is the mean true redshift of the frequency band for the line $j$.\footnote{We use the same characteristic scales for all the volume probed by each line to reduce the computational complexity, with minimal impact in the resulting map unless the frequency band is very wide.} We apply Gaussian and top-hat filters in the transverse and radial direction, respectively, that in Fourier space take the form
\begin{equation}
    W(k,\mu) = \exp\left\lbrace\frac{-k^2\sigma_\perp^2(1-\mu^2)}{2}\right\rbrace{\rm sinc} \left(\frac{k\mu\sigma_\parallel}{2}\right)\,,
\end{equation}
where $k$ and $\mu$ are the modulus of the wavenumber and the cosine between its direction and its component along the line of sight, respectively.

We show an example of three-dimensional maps for the CO and {[CII]} lines observed at $z\in\left[2.5,3.5\right]$ in Fig.~\ref{fig:example_map}, for COMAP-like and FYST-like resolutions, respectively. We clearly observe the trade-off between angular and spectral resolution for each experimental configuration. Their combination provides complementary views of the underlying large-scale structure probed by line intensity mapping.

\subsubsection{Foregrounds and correlated continuum emissions}
\label{sec:foregrounds}
One of the key observational challenges for LIM experiments is the presence of foregrounds and continuum emissions that are correlated with the matter distribution. The dominant source of correlated contamination is the cosmic infrared background (CIB), which is the combined continuum IR emission from dust within galaxies (see, e.g., \citet{Serra:2014pva, Switzer:2018tel}). The CIB can be modelled by relating the SFR to the IR luminosity and assuming a dust spectral energy distribution (SED). The sources of contamination are therefore galaxies at a given redshift $z$ emitting into the target frequency $\nu$ of the LIM experiment. While implementing CIB emission is possible with the properties available within \SkyLine, for this work we rely on the results from \synsky for CIB maps and defer its implementation for future work. 

Another key source of contamination, particularly for radio frequencies, are Milky Way (MW) foregrounds. We include the option of adding MW emission using the Python Sky Model \textsc{PySM3} \citep{Thorne:2016ifb} in our simulations, with synchrotron, free-free, thermal dust, and anomalous microwave emissions as available components. Our fiducial model corresponds to the model~1 defined in the \textsc{PySM3} package for all components. The maps can then be smoothed, rotated to center a specific sky position, and added to the signal map.

\subsubsection{Instrument noise}\label{sec:intensity-noise}
After including all the sky signal (spectral lines plus continuum foregrounds) in the mock intensity maps, the instrumental noise is the only contribution left to add. We assume white noise with variance $\sigma_{\rm N}^2$ per voxel (or pixel), determined by the instrument sensitivity. The definition of the noise variance per pixel (voxel) for the angular (three-dimensional) maps depends on the kind of map used and on whether intensities or temperatures are used. For three-dimensional maps, the total noise variance per observed voxel, assuming only the autocorrelation between antennas is used, is
\begin{equation}
    \sigma_{{\rm N}, I}^2=\frac{\sigma_{\rm pix}^2}{t_{\rm pix}'}\,, \qquad   \sigma_{{\rm N},T}^2 = \frac{T_{\rm sys}^2}{\delta\nu t_{\rm pix}'}\,,
    \label{eq:sigmaN_T}
\end{equation}
for intensities and temperatures, respectively, where $\sigma_{\rm pix}$ is the noise equivalent intensity, NEI (see e.g., Table 1 in \citet{CCAT-Prime:2021lly}), $T_{\rm sys}$ is the effective system temperature, and $t_{\rm pix}$ is the total effective observing time per pixel; $t'$ is equivalent to the product of the total number of detectors observing and the observing time per pixel. Then, for each voxel or pixel, we sample a Gaussian distribution with zero mean and variance $\sigma_{\rm N}^2$, rescaled accordingly in case $N_{\rm ss,\theta},N_{\rm ss,\nu}>1$. 

\subsection{Galaxy number counts}
\label{sec:gals}
Cross-correlations of line-intensity maps with galaxy catalogs are a promising probe of both cosmology and astrophysics. Indeed, they have led to the first claimed detections of cosmological signals in the clustering regime at large scales (see e.g.,~\citet{Pullen:2017ogs, Yang:2019eoj, CHIME:2022kvg, Cunnington:2022uzo}). It's of significant interest to include the signal of galaxy -- line cross-correlations and we outline below how this can be achieved within the context of our framework. \par 
\universemachine predicts a stellar mass and SFR for every subhalo in our catalog. Defining the specific star formation rate as ${\rm sSFR} = {\rm SFR}/M_*$, the distribution of subhalos in the $M_*$--sSFR plane can be seen to be broadly bimodal \citep{2009ARA&A..47..159B}, leading to a population of star-forming and a population of quenched galaxies. Observationally, these populations broadly correspond to, e.g., luminous red galaxies (LRGs) for the quenched population, and emission line galaxies (ELG) for the star-forming population. \par
Since cross-correlating these galaxy populations with line-intensity maps is of observational interest, we use \universemachine to also produce galaxy number count mocks. Quenched (``LRG'') and star-forming (``ELG'') populations are selected by performing cuts in sSFR, and a desired number density is achieved by adopting a $M_*$ cut. The galaxy catalogs are then converted to maps by converting their angular coordinates and redshifts to Cartesian coordinates and interpolating them to a grid, similarly to their line intensities. We demonstrate the cross-correlations between our simulated galaxy and line intensity maps in \S~\ref{sec:cross}.
\begin{figure*}
    \centering
    \includegraphics[width=\textwidth]{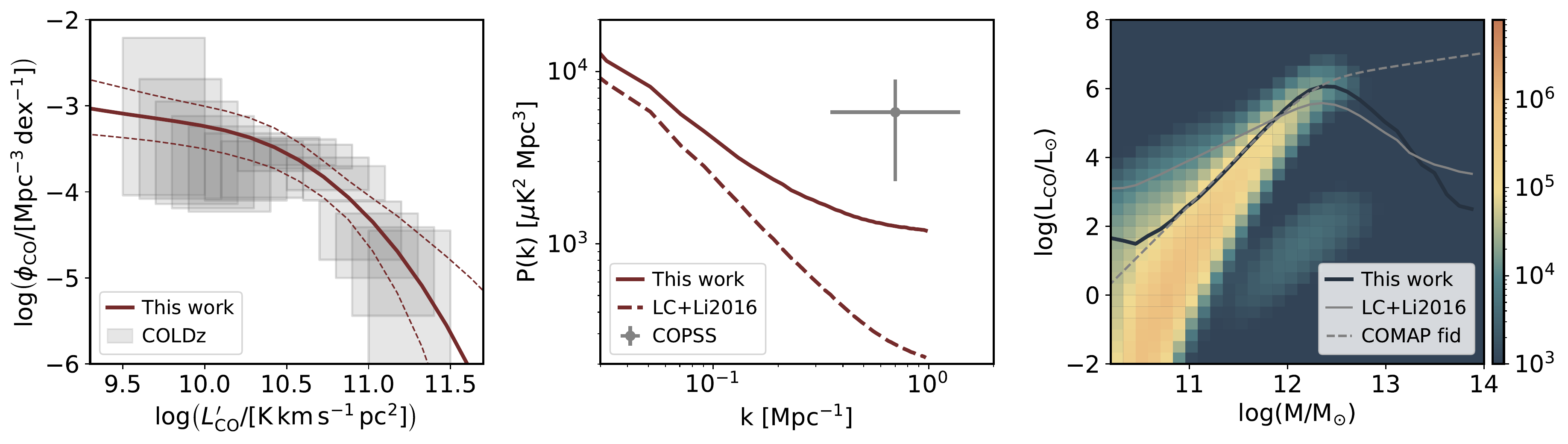}
    \caption{\SkyLine fit to the COLDz luminosity function. The leftmost panel shows the CO luminosity function measured by COLDz at $z=2-2.7$, with the $2\sigma$ error boxes shown in grey, and the luminosity function measured from \SkyLine using the best-fit values for the $L_{\rm IR}$-to-$L'_{\rm CO}$ relation, with the corresponding $1\sigma$ envelopes in dashed lines. The central panel shows the CO power spectrum measured by COPSS at $z=2.8$ and the predicted power spectra from our scheme, using our best-fit model as well as the model of \protect\cite{Li:2015gqa}. In the rightmost panel, we show the 2D histogram for the CO luminosity-to-halo mass distribution using our best-fit model. We compare our fiducial result to the prediction when the model by \protect\cite{Li:2015gqa} is adopted, as well as the fiducial COMAP model. }
    \label{fig:lf_function}
\end{figure*}

\subsection{Summary statistics}
\label{sec:sumstat}
We consider two key summary statistics that can be measured from LIM maps: multipoles of the anisotropic power spectrum, $P_\ell (k)$, and the voxel intensity distribution (VID), $\mathcal{B}_i$, although any summary statistic can be computed from the resulting mock maps, such as higher-order correlations (see e.g., \citet{Beane:2018pmx}), the deconvolved distribution estimator \citep{Breysse:2022fdi, Chung:2022zeu} or antisymmetric cross-correlations (see e.g.,~\citet{Sato-Polito:2020qpc}). The three-dimensional anisotropic power spectrum $P(k,\mu)$ can be decomposed into Legendre multipoles
\begin{equation}
    P_{\ell}(k) = \frac{2\ell+1}{2}\int_{-1}^{1} d\mu\ P(k,\mu) \mathcal{L}_{\ell}(\mu),
\end{equation}
which we compute using the \textsc{nbodykit}, and where $\mathcal{L}_{\ell}(\mu)$ are the Legendre polynomials. We focus our results on the monopole and quadrupole terms of the expansion. 

Another alternative, given a realization of an angular map, is to compute the angular power spectrum 

\begin{equation}
    C_{\ell}^{XY} = \frac{1}{2\ell + 1} \sum_{m=-\ell}^\ell a_{\ell m}^X a_{\ell m}^{Y*}\,,
\end{equation}
where $a_{\ell m}^X$ are the spherical harmonic coefficients of a map $X$. These are computed using \textsc{healpy}. 

We further analyze the simulated LIM maps by computing the VID, which is an estimator for the 1-point probability distribution function $\mathcal{P}(I)$ of the intensity within a voxel, and is defined as the number of voxels (or pixels) $\mathcal{B}_i$ within a given intensity bin $\Delta I_i$:
\begin{equation}
    \mathcal{B}_i = N_{\rm vox}\int_{\Delta I_i}dI\mathcal{P}(I)\,,
\end{equation}
$N_{\rm vox}$ is the total number of voxels in the observed volume. In practice, we directly measure $\mathcal{B}_i$ from our maps by computing a histogram. The covariance between the power spectrum and the VID depends on the integrated bispectrum~\citep{Sato-Polito:2022fkd}, which could also be measured from the map.

While not directly observable with LIM experiments, we also consider the line-luminosity function $\phi_j(L)$, defined as the number of objects per unit comoving volume per luminosity. We compute the luminosity function from our lightcones in order to compare it with additional data sets and calibrate relations between line luminosities and other modeled quantities.

\section{Analysis and Results}
\label{sec:analysis}
\subsection{Spectral lines and noise}
\label{subsec:lines_and_noise}

To showcase the capabilities of our lightcone simulation, we choose a handful of spectral lines to target and observational specifications to model. While there are a variety of ongoing and planned LIM experiments, we define a standardized instrument across all spectral lines. We choose to do so in order to compare the signal of each line emission on equal footing \textemdash at the same redshifts, across the same angular scales, and with the same sensitivities. We consider emissions from CO, {[CII]}, and both Ly$\alpha$ and HI, and discuss the specific astrophysical properties of each line in the subsections below.

Our fiducial maps cover a redshift range between $z=2.5-3.5$ with a survey area of $\Omega_{\rm field} = 400$~deg$^2$. The spectral resolution is fixed to $R=\nu_{\rm obs}/\delta_{\nu}={220}$ and the angular resolution is $\theta_{\rm FWHM} = 2'$. We select the instrument noise such that the power spectrum has a signal-to-noise ratio of 5 at $k=0.1$~Mpc$^{-1}$, which results in a higher noise for the brighter lines. Finally, since LIM experiments are expected to not be able to measure absolute intensity zero-points (due to instrumental limitations and the removal of continuum foregrounds), we remove the mean from all the simulated maps. 

\subsubsection{CO}
The CO luminosity is usually expressed in terms of pseudo-luminosity $L'$ in K km/s pc$^2$ units~\citep{2016ApJ...829...93K}, which is given by
\begin{equation}
    \frac{L_{\rm CO(J\rightarrow J-1)}}{L_\odot} = 4.9\times 10^{-5}J^3 \frac{L'_{\rm CO(J\rightarrow J-1)}}{{\rm K\ km\ s^{-1}\ pc^2 }\ L_\odot^{-1}}\,.
\end{equation}
We therefore model the CO luminosity using Eq.~\ref{eq:LtoLIR}, using $L'_{\rm CO(J\rightarrow J-1)}$ instead in the left-hand side. This is the same approach adopted by \citet{Li:2015gqa}, but since we have updated the underlying galaxy formation model (resulting in a different distribution of SFR as function of halo properties) and SFR-to-$L_{\rm IR}$ relation, we also update the coefficients of Eq.~\ref{eq:LtoLIR} as described in \S~\ref{sec:pops_maps}. We use $\alpha^{\rm CO}_{\rm IR} = 1.2$ 
and $\beta^{\rm CO}_{\rm IR} = -4.2$, obtained from a fit to observational data as described in \S~\ref{sec:pops_maps}. 

Notice that by assigning a CO luminosity to each halo based on its SFR, the halo-to-halo scatter in the SFR from \textsc{UniverseMachine} is naturally captured, but the scatter in the $L_{\rm CO}$-to-SFR relation must still be added. We therefore add a log-normal scatter to $L_{\rm CO}$, parametrized by $\sigma_{L_{\rm CO}} = 0.2$.

\subsubsection{{[CII]}}
We model the {[CII]} emission using the parameterization introduced in \citet{2018A&A...609A.130L}, in which the line luminosity is related to the halo SFR through Eq.~\eqref{eq:SFRtoL_PL}. The parameters are assumed to be redshift dependent and given by
\begin{equation}
    \alpha^{\rm CII}_{\rm SFR} = 1.4-0.07z, \quad \text{and} \quad \beta^{\rm CII}_{\rm SFR} = 7.1-0.07z,
\end{equation}
with a log-normal scatter in the $L_{\rm CII}$-SFR relation of $\sigma_{L_{\rm CII}} = 0.5$.

\begin{figure*}
    \centering
    \includegraphics[width=0.95\linewidth]{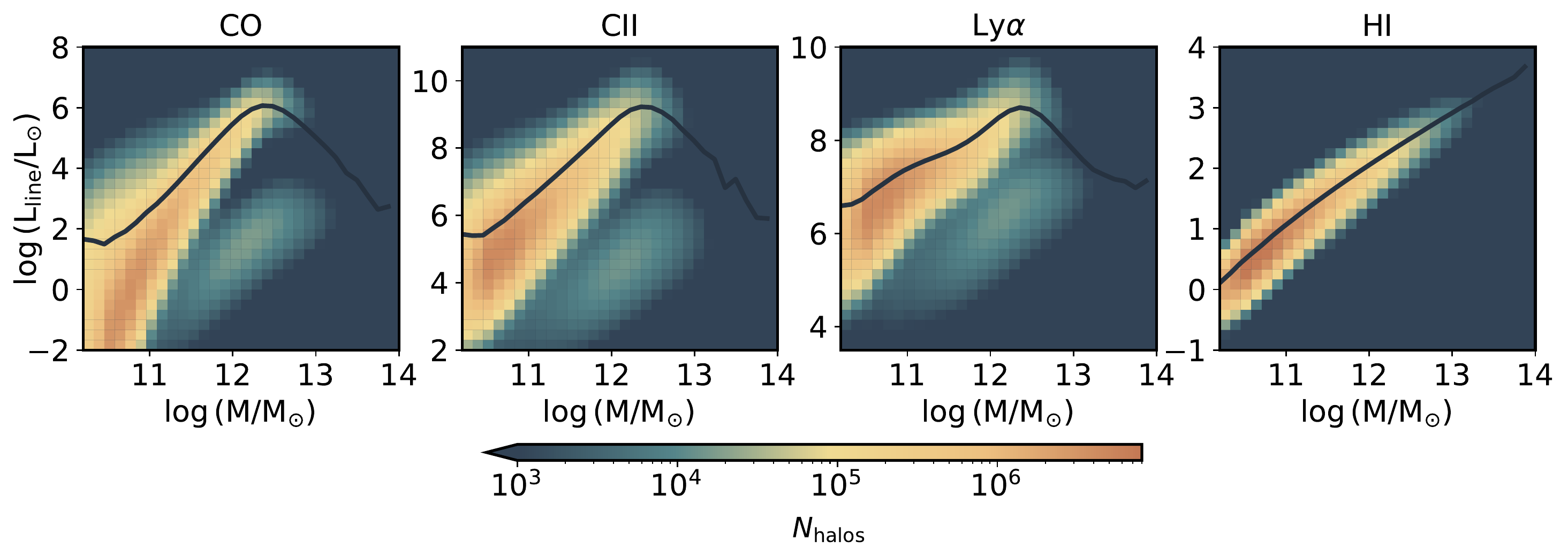}
    \caption{Two-dimensional histograms  of line luminosities as a function of halo mass for redshifts between 2.5 and 3.5. The solid lines correspond to the mean luminosity in each halo mass bin.}
    \label{fig:LofM}
\end{figure*}

\subsubsection{Ly$\alpha$}
The Ly$\alpha$ line can be modelled using an intrinsic SFR-to-$L_{\rm Ly \alpha}$ relation and an escape fraction that captures the scattering and attenuation of the Ly$\alpha$ luminosity. Though largely unconstrained, there are two limiting trends derived from observations: $f_{\rm esc}$ grows monotonically with redshifts and decreases with higher SFR. An effective parameterization following these trends and calibrated to Ly$\alpha$ emitters was derived by~\citet{COMAP:2018svn}:
\begin{equation}
\begin{split}
    f_{\rm esc}&({\rm SFR}, z) = \left(1+e^{-1.6z+5}\right)^{-1/2} \times\\ & \times\left[0.18 + \frac{0.82}{1+0.8\left(\frac{\rm SFR}{M_{\odot} \text{yr}^{-1}}\right)^{0.875}}\right]^2\,.
\end{split}
\end{equation}
Then, if we add that to known ratios between Ly$\alpha$ and other lines and use the Kennicutt-Schmidt relation, we find~\citep{COMAP:2018svn}
\begin{equation}
    \frac{L_{\rm Ly\alpha}}{L_\odot} = 1.6\times 10^{42} \left(\frac{\rm SFR}{M_{\odot} \text{yr}^{-1}}\right) f_{\rm esc}({\rm SFR}, z)\,.
\end{equation}

\subsubsection{HI}
Finally, the model of post-reionization HI luminosity is inherently different than higher frequency lines. 
In the post-reionization universe, the majority of the neutral hydrogen mass $M_{\rm HI}$ is bound in dark-matter halos. We connect $M_{\rm HI}$ directly to the mass $M$ of the host halo using the results based on the TNG100 magneto-hydrodynamic simulation \citep{Villaescusa-Navarro:2018vsg}. Their results are well modelled as a power-law suppressed at lower mass given by Eq.~\ref{eq:LofM_HI}, with $\beta_M = 0.35$ and the remaining parameters given by the best-fit values for FoF halos in their Table~1. The luminosity produced by a single halo with mass $M$ is then given by
\begin{equation}
    L_{\rm HI} = \frac{3 A_{10} h \nu_{21}}{4 m_p} M_{\rm HI}(M),
    \label{eqn:lum_HI}
\end{equation}
where $M_{\rm HI}(M)$ is given by Eqn.~\ref{eq:LofM_HI}, $h$ is the Planck constant, $m_p$ is the proton mass, and  $A_{10}\simeq 2.869 \times 10^{-15}$s$^{-1}$ is the spontaneous emission coefficient (see, e.g., \citet{Bull:2014rha}).
\begin{figure*}
    \centering
    \includegraphics[width=\textwidth]{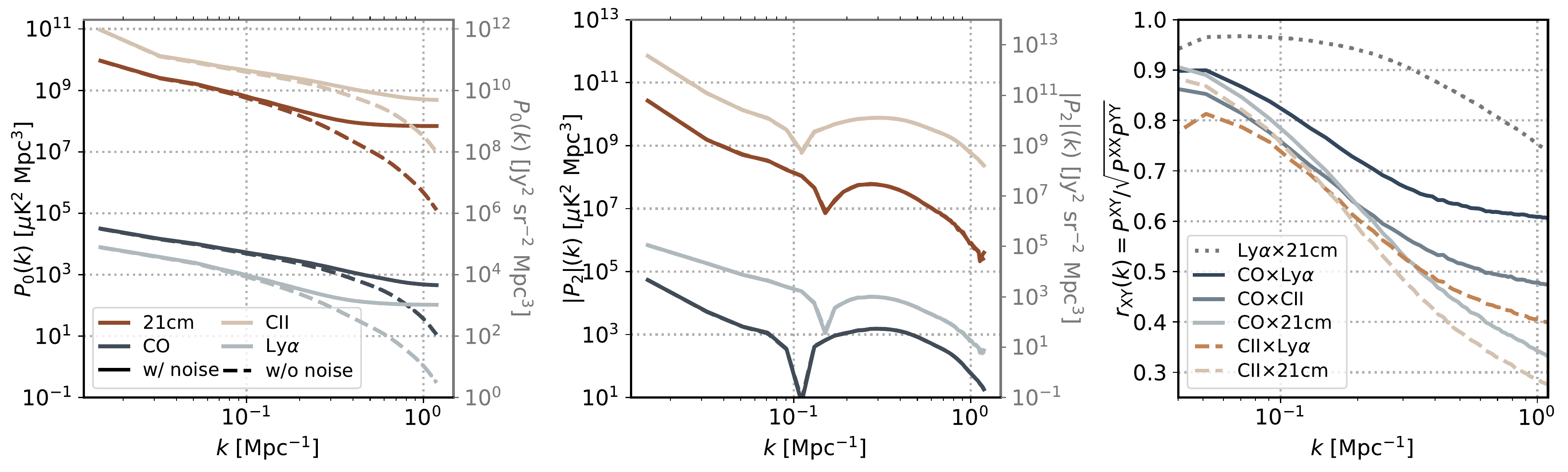}
    \caption{Power spectrum monopole (left) and quadrupole (center, showing its absolute value, where the values at $k$ higher than dip are negative) measured from the simulated maps using fiducial instrumental parameters. Following standard conventions in the literature, CO and HI are computed in brightness temperature units with values shown on the left axes, whereas {[CII]} and Ly$\alpha$ are computed as intensities and correspond to the right axes. Solid and dashed curves correspond to simulations with and without instrument noise, respectively. The right panel shows the cross-correlation coefficient of the power spectrum monopole between CO(1$\rightarrow$0), [CII],  Ly$\alpha$, and HI lines. In all cases the noise is set to achieve a signal-to-noise ratio of the power-spectrum monopole equal to 5 at $k=0.1\,{\rm Mpc^{-1}}$ and $z\in\left[2.5,3.5\right]$ is considered.}
    \label{fig:Pk}
\end{figure*}

\subsection{Populations and maps}
\label{sec:pops_maps}
A variety of approaches have been employed to model the CO intensity fluctuations by relating halo properties to line luminosity. For example, \citet{Li:2015gqa} adopts an empirically-driven model in order to relate SFR and $L_{\rm CO}$. Since SFR is not a directly observable quantity, this relation relies on an intermediate tracer of star formation, which \citet{Li:2015gqa} take to be the IR luminosity, which shows a higher correlation with $L_{\rm CO}$ than the SFR. \citet{Chung:2021vzh} and \citet{Padmanabhan:2017ate} instead choose to directly parametrize the relation between $L_{\rm CO}$ and halo mass $M$ using double power-laws, thereby compressing many parameters present in the model by \citet{Li:2015gqa} that are degenerate in a LIM experiment.

We update the parameters in the SFR-to-$L'$ relation by fitting the luminosity function derived from our lightcones to the measurement by the CO Luminosity Density at High Redshift survey (COLDz; \citet{Riechers:2018zjg}) at redshift $z\sim 2.4$. \citet{COMAP:2021rny} performed a similar analysis for a double-power law relating $L_{\rm CO(1-0)}$ directly to the halo mass. We consider the combined fields of COSMOS and GOODS-N, and include only the 4 independent luminosity bins for the CO($1\rightarrow 0$) transition in our fit.  We find $\alpha^{\rm CO}_{\rm IR} = 1.2 \pm 0.3$ and $\beta^{\rm CO}_{\rm IR} = -4.2\pm 4$. Note  that \citet{Li:2015gqa} adopts a different definition of the coefficients. The equivalent parameter values can be found by a simple change of variables, which result in $\alpha'^{\rm CO}_{\rm IR} = 0.8 \pm 0.2$ and $\beta'^{\rm CO}_{\rm IR}=3.4 \pm 2.4$. The preferred $\alpha'^{\rm CO}_{\rm IR}$ and $\beta'^{\rm CO}_{\rm IR}$ values are very correlated, following the relation
\begin{equation}
    \alpha'^{\rm CO}_{\rm IR} = -0.094\beta'^{\rm CO}_{\rm IR} + 1.14\,.
\end{equation}
We show, in the first panel of Fig.~\ref{fig:lf_function}, the fiducial CO(1-0) luminosity function with the 1$\sigma$ uncertainty envelope, along with the 2$\sigma$ error boxes given by COLDz. We compare the CO power spectrum measured by the reanalysis of COPSS~\citep{Keating:2020wlx} at $z\sim 2.8$ with the \SkyLine results. We find better agreement between the model presented here and the COPSS power spectrum measurement than the previous model by \citet{Li:2015gqa}. However, COPSS measured shot noise power spectrum is still $\sim 2\sigma$ higher than our prediction. This may be due to an underestimation of the effects of cosmic variance: the shot noise amplitude depends on the second moment of the luminosity function, which is dominated by the brightest sources and therefore for small volumes like COPSS it is subject to a large cosmic variance~\citep{Gkogkou:2022bzo}. Finally, we compare the relation between CO luminosity and halo mass derived in our work with the relation found by \citet{COMAP:2021rny} ---which has become the fiducial model for COMAP--- and the model by \citet{Li:2015gqa}. The main discrepancy found between our results and those by COMAP lie in the poorly-constrained quenched massive halos, and it has a marginal impact in the predictions.

We also compute the line-luminosity functions for the remaining spectral lines considered in these examples. We show in Fig.~\ref{fig:LofM} the predicted distribution of line luminosity as a function of halo mass from \textsc{UniverseMachine} coupled to empirical relations between spectral line-emissions and SFR. Our results capture the full distribution of luminosities, which includes quenched and star-forming galaxy populations.

\subsection{Auto- and cross-power spectra}
We compute power spectra for each of the line-intensity maps considered and the cross-power spectra between them, without including contamination. We show in Fig.~\ref{fig:Pk} the monopole and quadrupole of the power spectrum, with and without instrument noise, for each of the modeled spectral lines using the survey designs described in \S~\ref{subsec:lines_and_noise}. These results capture a variety of observational effects in the clustering statistics of LIM, including the spectral and angular resolutions, survey geometry, redshift evolution within the lightcone, and redshift-space distortions. We can see the that the instrument noise flattens the power spectrum monopole at small scales, whereas the noiseless map has diminishing power due to the effect of smoothing. Since the instrumental noise is isotropic, it does not contribute to the quadrupole of the power spectra. Note that the quadrupole is largely affected by the anisotropic resolution limit at scales comparable to
$\sigma_\perp$ and $\sigma_{\parallel}$, and the zero-crossing for each line takes place at different values of $k$ due to differences in the line-luminosity bias (see e.g., \citet{Chung:2019iim,Bernal:2019jdo}).

Given the multi-tracer nature of upcoming LIM observations and the complementarity between the astrophysical information encoded in various line emissions, understanding the correlation between different lines will be key~\citep{Sun:2020mco,Schaan:2021gzb}. We compute the cross-correlation coefficient 
\begin{equation}
    r_{\rm XY}(k) = \frac{P^{\rm XY}_0(k)}{\sqrt{P^{\rm XX}_0(k) P^{\rm YY}_0(k)}},
\end{equation}
between each line X,Y $\in$ \{CO, {[CII]}, Ly$\alpha$, HI\}, and display them in the right panel of Fig.~\ref{fig:Pk}. 

In the large-scale 2-halo dominated regime of the power spectrum, the lines are expected to be highly correlated, although white noise at large scales such as instrumental and shot noise, as well as higher order bias\footnote{We might expect higher order biases like $b_2$ also cause mild decorrelation at the largest scales. The spectrum $\langle \delta^2 \delta^2 \rangle$ is degenerate with white noise in the $k\to0$ limit, but is present in the cross-correlation as well as the auto.} prevent $r_{\rm XY}$ from being exactly unity. On smaller scales, the power spectrum receives contributions from the 1-halo and shot noise terms, which decorrelate the lines. Ly$\alpha$ and HI are more correlated than the rest because both lines are related to neutral hydrogen, hence they trace somewhat similar halos, while the other lines hold slightly larger differences and present larger scatter. 

\begin{figure*}
    \centering
    \includegraphics[width=0.95\textwidth]{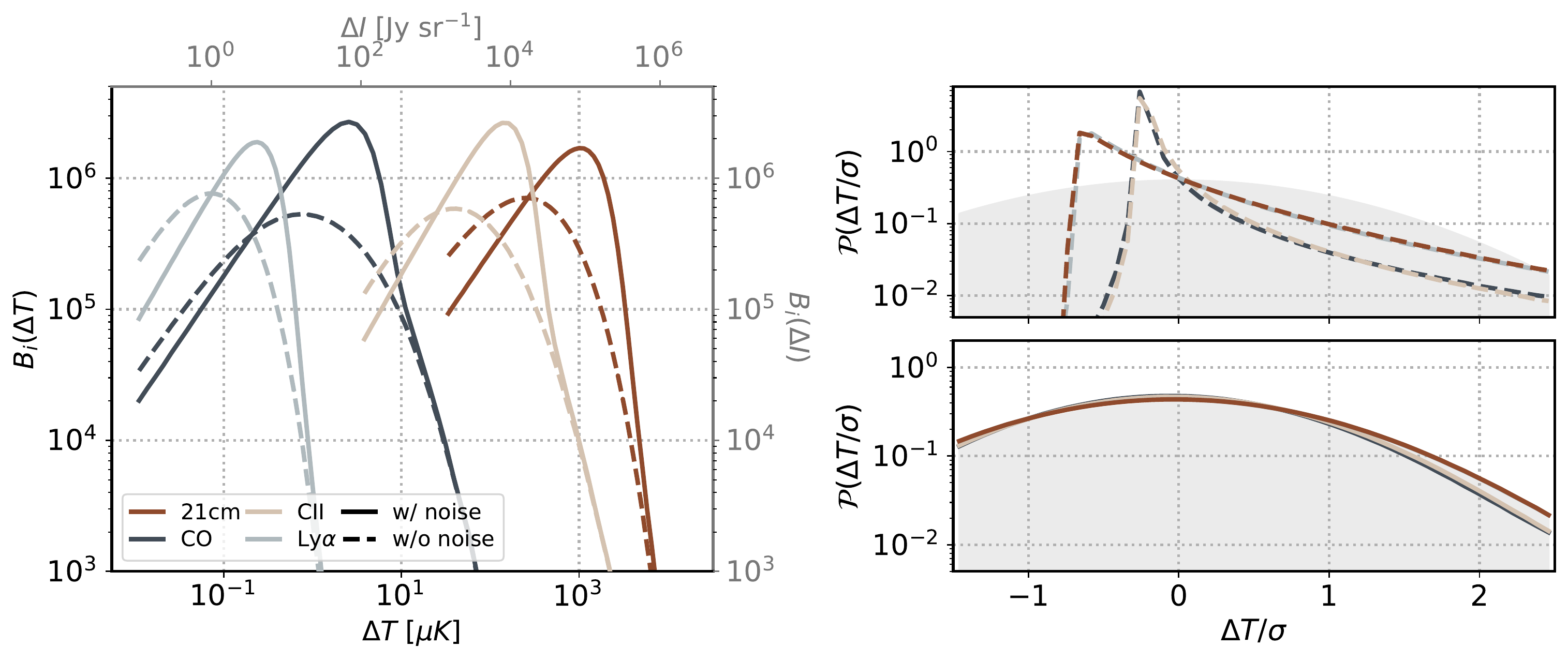}
    \caption{Left: Voxel intensity distribution for CO($1\rightarrow 0$), [CII], Ly$\alpha$, and 21cm lines  measured from simulated maps using the fiducial instrument design. Similarly to Fig.~\ref{fig:Pk}, CO and 21cm are in brightness temperature units and correspond to the bottom axis, while {[CII]} and Ly$\alpha$ are computed as intensities and correspond to the top axis, with solid and dashed lines denoting maps with and without instrument noise. Right: Normalized histograms in studentized units with (bottom) and without noise (top), with the noise distribution in gray shade.}
    \label{fig:VID}
\end{figure*}

\subsection{Voxel intensity distribution}
As stated in \S~\ref{sec:sumstat}, the VID is an estimator of the one-point probability distribution function of intensities measured in a voxel, hence it depends on the size of the voxel. Furthermore, there is a trade-off between small voxels -- to include maximal information -- and large voxels -- to reduce pixel-to-pixel correlations and the instrumental noise. We follow~\citet{Vernstrom:2013vva}, which found optimal to choose $\theta_{\rm FWHM}$ as the pixel size, and use the channel width $\delta \nu$ as the size of the voxel in the direction along the line of sight for three-dimensional maps. 

We show the VID, with and without instrumental noise and using a logarithmic binning in the intensity or temperature, for the lines considered in this work in the left panel of Fig.~\ref{fig:VID}. Note that the difference along the $x$-axis is due to the width of the distribution, which after mean-subtraction has its mean at $\Delta T=0$. Low intensities are totally dominated by instrumental noise once it is added to the map, but the high intensity tail still contains information from the intrinsic signal. In the right-hand panels we show normalized histograms for each simulated maps using studentized units (i.e., dividing the temperature or intensity of each voxel by the standard deviation of their distribution).  

While the intensity mapping signal is inherently integrated across all emitters, our simulated maps can be a valuable tool to break down the contribution from different source populations. In the case of the VID, we study the properties of halos that contribute to voxels of a given temperature, thereby gaining a deeper understanding of the information captured by this summary statistic. We show a particular example here, but leave a broader and more systematic study for future work.

\begin{figure}
    \centering
\includegraphics[width=0.8\linewidth]{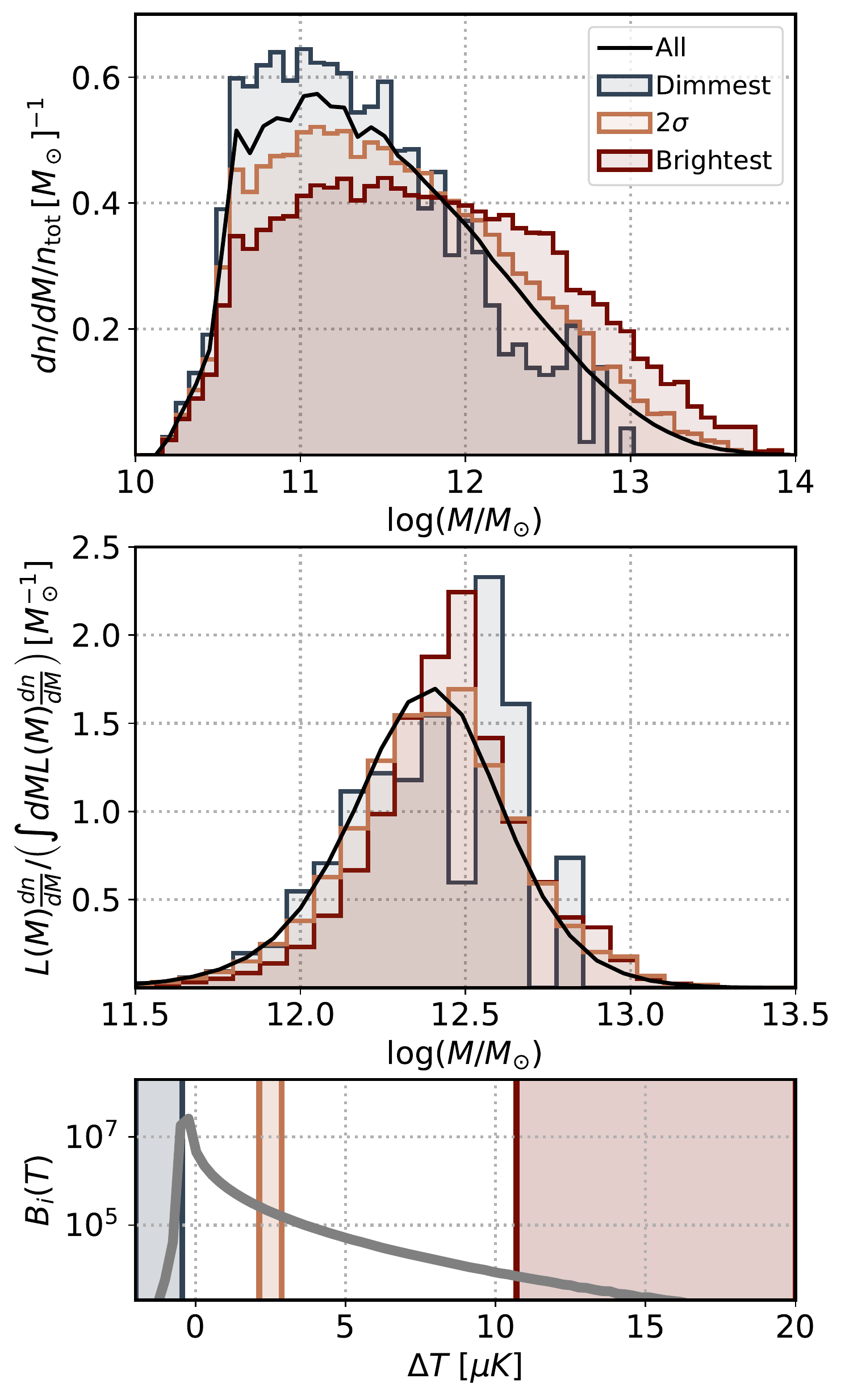}
    \caption{Normalized halo mass function (top) and luminosity-weighted halo mass function (middle) of  halos residing in specific subsets of voxels for the CO($1\rightarrow 0$) case described in the main text. We show results for the whole sample (black) and for three subsamples containing 1\% of the total number of voxels each: the dimmest voxels (blue), the brightest voxels (red) and the voxels with temperatures around 2 times the standard deviation over the mean (orange), as shown with the respective colors in the VID of the map (bottom).}
    \label{fig:voxel_LF}
\end{figure}

Let us consider our case study for CO(1-0), the VID of which we show in the bottom panel of Fig.~\ref{fig:voxel_LF} using linear bins in $\Delta T$. We sub-select voxels by their brightness temperature and compute the normalized halo mass function and the line-luminosity-weighted mass function of the halos inhabiting such voxels (top panels of Fig.~\ref{fig:voxel_LF}). Note that to account for the logarithmic binning in halo mass we also weight the histograms by halo mass, in addition to by the luminosity when needed. Specifically, we select three sets of voxels that contain 1\% of the total number of voxels each: the brightest voxels, the dimmest voxels, and those which have temperatures around 2 times the standard deviation over the mean ($\sim 2.5\,\mu{\rm K}$ for our example). 
This allows us to compare the population of halos that contribute to faint or bright voxels, respectively. 

This preliminary example demonstrates there is no one-to-one relationship between the mass of the halos and the brightness of a voxel. All voxels have a very wide mass function, with the dimmest voxels more dominated by lighter halos than the brightest voxels. In general, as a voxel is brighter the mass distribution of the halos within flattens. This clear trend disappears once we account for the luminosity of each halo (which is what determines the contribution to the voxel intensity, see Eq.~\eqref{eq:LtoI}). We additionally find a bimodal luminosity-weighted halo mass distribution for the dimmest voxels, indicating a sizable population of relatively isolated bright and heavy emitters, while the distribution for the brightest voxels peaks at intermediate masses. These results broadly align with the $L(M)$ relation shown in Fig.~\ref{fig:LofM}, but also show that there are voxels hosting a wide range emitters, especially for the brightest voxels. The brightest emitters, even in isolation, surpass the intensity of many faint emitters that constitute the faintest voxels. In this example, where the VID is very peaked, the whole sample is dominated by voxels at the mean temperature (i.e., $\Delta T = 0$). The properties of emitters near the mean are very similar to the whole sample, and we do not include it in the figure to ease its reading.

\subsection{Line interlopers}
Even after removing continuum contributions to the intensity maps, emission lines from different volumes redshifted to the same observed frequencies remain in the measurements. Fortunately, if modeled suitably, these contributions are not contaminants but a source of astrophysical and cosmological information, even if they may dominate the signal (see e.g.,~\citet{Gong:2020lim}). 

\begin{figure}
\centering   
\includegraphics[width=\linewidth]{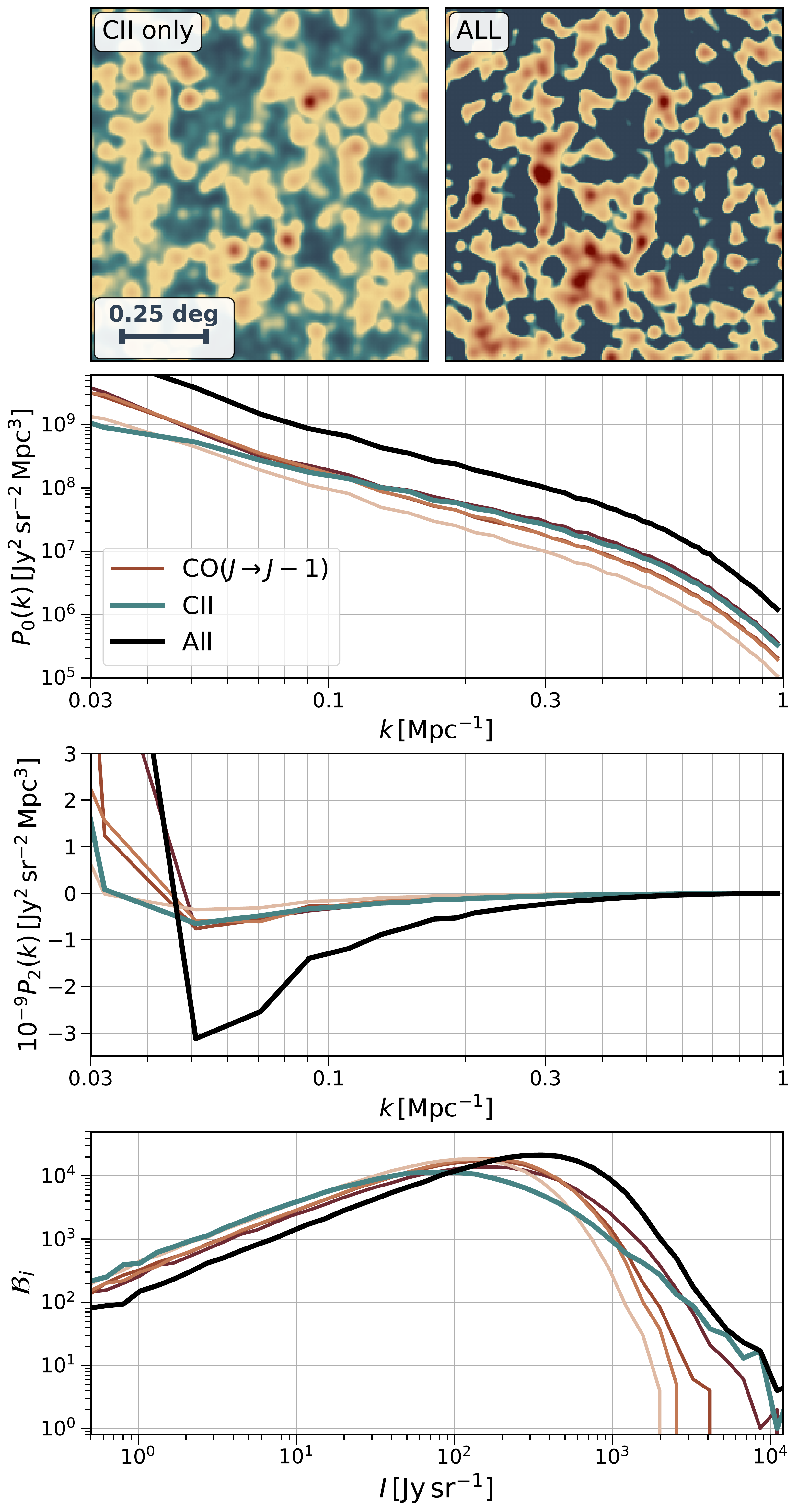}
    \caption{Zoomed 1 deg$^2$ projected simulated map including only the contribution from {[CII]} at $z=5$, within a bin of $\Delta z = 1$, (top left) and also including interloper contributions from CO rotational transitions from $J=4$ to $J=7$ (top right). Bottom panels show the power spectrum monopole and quadrupole, and the VID (from top to bottom) for each independent line (lighter shades of red denote higher values of $J$) and the map with all the lines included for the whole 36 deg$^2$ considered.}
    \label{fig:interlopers}
\end{figure}

As mentioned in Section~\ref{sec:map_lines}, \SkyLine can include line interlopers in the mock maps. We show an example in Fig.~\ref{fig:interlopers}, where we compare the maps and summary statistics of a single target line and when interlopers are included. We focus on {[CII]} at $z=5$, for a redshift bin $\Delta z = 1$, which corresponds to observed frequencies $\nu_{\rm obs}\in \left[292.4,345.6\right]\, {\rm GHz}$. This frequency band also receives contributions from CO rotational transitions from $J=4$ to $J=7$; lower transitions fall below the observing frequency band, and transitions from $J\geq 8$ become negligible. We use results from~\citet{2016ApJ...829...93K} to relate the intensity from each transition to CO(1-0), with ratios $L'_{{\rm CO}(J\rightarrow J-1)}/L'_{{\rm CO}(1\rightarrow 0)}=r_J$ given by 
$r_4=0.57$, $r_5=0.32$, $r_6=0.19$ and $r_7=0.1$.\footnote{\citet{2016ApJ...829...93K} obtained these ratios from low-redshift observations by \textit{Herschel}/SPIRE, finding $\alpha_{{\rm CO}(1\rightarrow 0)}=1.27$ and $\beta_{{\rm CO}(1\rightarrow 0)}=-1$. We apply the same ratios at higher redshifts and to the values of $\alpha_{{\rm CO}(1\rightarrow 0)}$ and $\beta_{{\rm CO}(1\rightarrow 0)}$ obtained in Section~\ref{sec:pops_maps}. A precise determination of the parameters controlling the scaling relations for transitions from $J>1$ at these redshifts is left for future work.}

Even if {[CII]} is the brightest emission line in star-forming galaxies in the far IR and lower frequencies, the contribution from the foreground interloper CO lines 
($z=\left\lbrace 0.46,\, 0.82,\, 1.18,\, 1.55\right\rbrace$ for $J$ growing from 4 to 7, respectively)  dominate the signal due to higher intensities (due to the lower redshift) and the projection effects (which boost their measured clustering and introduce strong anisotropies; see e.g.,~\citet{Lidz:2016lub}). 
The projection along the lightcone of the interloper signal to the volume where the target signal comes from also introduces strong artificial anisotropies in the line-intensity map. The same angular (frequency) scale corresponds to smaller  (larger) physical scales for the interloper than for the target spectral line, but they are all interpreted at the same $k$ values due to the redshift confusion. This can be seen in the middle panels of Fig.~\ref{fig:interlopers} where the CO power spectrum monopoles have different slopes, and where the power spectrum quadrupoles change significantly with respect to the {[CII]}-only case due to the large artificial anisotropy introduced. Comprehensive accounting of all interlopers allows for more coherent astrophysical modeling, and enables the ability to probe several redshifts with the same set of observations.

\subsection{Galactic foregrounds}
\label{sec:foregrounds}
To illustrate the effect of foregrounds on the summary statistics of interest, we add the MW components described in \S~\ref{sec:foregrounds}. We focus on the fiducial CO(1-0) model and fiducial observational parameters ($z=2.5-3.5$, $R$=700, and $\theta_{\rm FWHM} = 2'$), but with a restricted survey area of 4~deg$^2$. We generate foreground maps for each frequency bin, smoothed with the appropriate angular resolution of the instrument, and rotate the map to center it on the observed field, which we choose in this simple example to be the galactic south pole. We then project the foreground \textsc{HEALpix} maps using the flat-sky approximation. We do not include any additional sources of contaminant intensity, such as line interlopers or instrument noise, since our goal is to demonstrate the impact of the MW on LIM observables. 

\begin{figure}
\centering   
\includegraphics[width=0.7\linewidth]{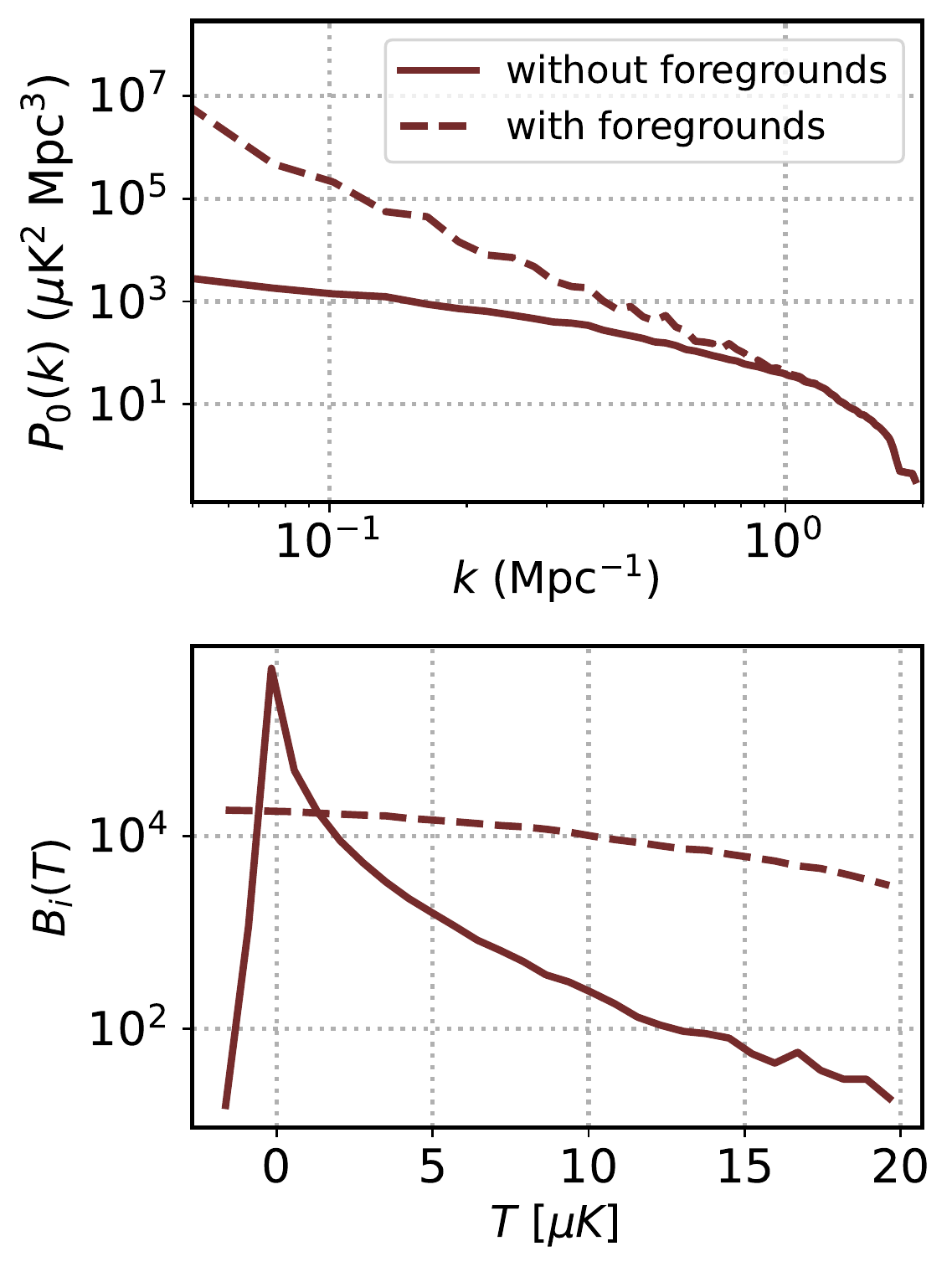}
    \caption{Power spectrum (top) and VID (bottom) with (dashed) and without (solid) the contribution from Galactic foregrounds without any mitigation. We include the signal from CO(1-0) emission in a 4 deg$^2$ lightcone from $z=2.5-3.5$ and the contribution from foregrounds at the corresponding frequencies of $\nu_{\rm obs}=26-33$~GHz.}
    \label{fig:foreground}
\end{figure}

We show the CO power spectrum and the VID with and without the contribution from Galactic foregrounds in Fig.~\ref{fig:foreground}. As expected, without any mitigation technique, the signal is completely dominated by the foregrounds. The contribution from foregrounds significantly grows at large scales following an approximate $k^{-3}-k^{-4}$ dependence, while at small scales the power spectrum with and without foregrounds coincide due to its smaller contribution and the resolution cut off. On the other hand, the VID with foregrounds is approximately flat for the range of temperatures of the signal-only VID because the foreground VID extends to significantly higher temperatures.

We stress, however, that this result will be highly dependent on the particular survey footprint. Our choice of a 4~deg$^2$ square survey on the galactic south pole may not accurately represent the level of foreground emission expected for any ongoing experiment, but rather serves to demonstrate the potential of \SkyLine as platform to develop and test foreground mitigation techniques.

\section{Cross correlations with other probes}
\label{sec:cross}
Despite LIM being a nascent field, forthcoming LIM surveys will overlap with other experiments and missions over the same regions of the sky~\citep{Bernal:2022jap}, which has spurred a significant interest in cross-correlating line-intensity maps with other observables. The first detections of most spectral lines have been achieved through cross-correlations with galaxies or quasar samples~\citep{Chang:2010jp, CHIMEdetection, Pullen:2017ogs, Yang:2019eoj, Kakuma:2019afo, Niemeyer:2022arn,Niemeyer:2022vrt, Cunnington:2022uzo}, as they boost the significance of LIM measurements and remove uncorrelated contaminants. Furthermore, cross-correlation with LIM can contribute to the reduction of photometric redshift uncertainties in galaxy surveys~\citep{Alonso:2017dgh,Cunnington:2018zxg,Modi:2021okf} and can be combined with weak-lensing surveys to constrain the impact of intrinsic alignments~\citep{Chung_2022}. The combination of LIM with cosmic microwave background secondary anisotropies can be used to improve the delensing of the primary anisotropies~\citep{Karkare:2019qla}, study reionization with high-redshift Sunyaev-Zel'dovich effect~\citep{LaPlante:2020nxx}, and to probe inflation with the late-time kinematic Sunyaev-Zel'dovich effect~\citep{Sato-Polito:2020cil}. This is a limited list of all the possibilities of combining LIM with other observables; see~\citet{Bernal:2022jap} for other opportunities. The amplitude and mutual information of many of these multi-probe observables can be readily quantified within our simulation framework. We demonstrate a few examples, with galaxy surveys, the CIB, and CMB secondary anisotropies below. 

\subsection{LIM $\times$ galaxies}
\label{sec:limxgals}
We first consider the cross-correlation between LIM measurements and galaxy distribution within the same volume. As an illustration, we consider two cases over a solid angle of 400 deg$^2$. First, we consider a low-redshift 21 cm survey over the redshift range $z\in \left[0.4,1.4 \right]$ in cross correlation with LRGs. This example approximately corresponds to a potential cross correlation between MeerKAT \citep{MeerKLASS:2017vgf} and DESI LRGs, hence we use a number density of LRGs of $n_{\rm LRG}=1.6\times 10^{-4}$ Mpc$^{-3}$ \citep{DESI:2022gle}. Second, we consider a higher-redshift example approximately following HETDEX \citep{Gebhardt:2021vfo}: a Ly$\alpha$ LIM survey over $z\in \left[2.2,3.2\right]$ cross-correlated with ELGs with number density $n_{\rm ELG}=1.0\times 10^{-4}$ Mpc$^{-3}$. In both cases we also consider an optimistic galaxy sample with 20 times more number density. We additionally note we do not consider the impact of flux limits on the redshift distribution $n(z)$ for now. More realistic galaxy mock samples involve flux cuts, rather than a cut in stellar mass for all the galaxies in the redshift bin under consideration. It is possible to implement such cuts within our framework, and we defer it to future studies assessing cross-correlations between galaxy and LIM surveys.

\begin{figure}
    \centering
    \includegraphics[width=\linewidth]{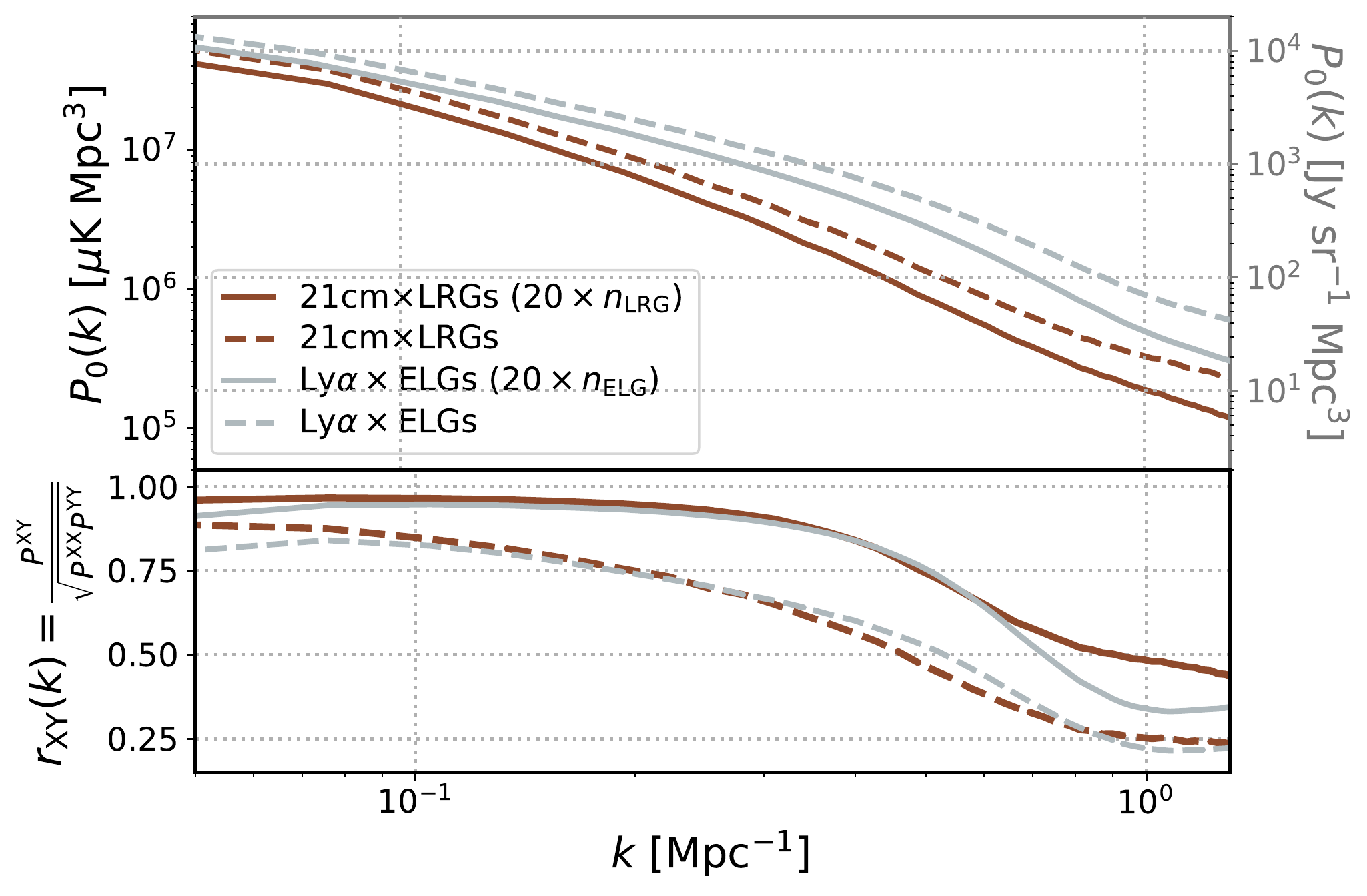}
    \caption{Cross-power spectrum monopole for 21cm and LRGs over $z\in\left[0.4,1.4\right]$ (red), and for Ly$\alpha$ and ELGs over $z\in \left[2.2,3.2\right]$ (grey), with their respective cross-correlation coefficients shown in the bottom panel. Both cases correspond to a solid angle of 400 deg$^2$ and we show a case in which the galaxy number density is 20 times larger (solid lines) than in the fiducial case (dashed) lines. Similarly to Fig.~\ref{fig:Pk}, 21cm is in brightness temperature units and Ly$\alpha$ is quantified as intensity, corresponding to the left and right axes, respectively.}
    \label{fig:cross_gal}
\end{figure}

We show these cross-power spectra in Fig.~\ref{fig:cross_gal}, along with the corresponding cross-correlation coefficients. As expected, the cases with lower number densities result in a higher amplitude of the power spectrum, because the galaxies included in the catalog correspond to higher $M_*$ (see \S~\ref{sec:gals}), thus more biased with respect to the matter density fluctuations. In turn, the cross-correlation coefficient shows the opposite trend: the more galaxies are included in the catalog, the higher the fraction of sources collected in the line-intensity map that are in both samples, hence the higher their correlation. However, note how the cross-correlation coefficient never reaches one, because of the difference in the shot noise between the two samples and their cross-power spectrum and the effects of nonlinear biases, as in Fig.~\ref{fig:Pk}. In both cases, the catalog with less galaxies starts to be less correlated with the LIM measurement at larger scales. 

\subsection{LIM $\times$ CMB secondary anisotropies and the CIB}
\label{sec:cross_CMB}

As mentioned previously, the default underlying lightcone of \SkyLine  was constructed in the same way as the simulated skies of \synsky \citep{Omori:SimSky}. The \synsky suite uses \textsc{MDPL2-UM} to model submillimeter wavelength extragalactic foregrounds to observations of the cosmic microwave background. These include several observables that generate CMB secondary anisotropies, including CMB lensing, thermal and kinetic Sunyaev-Zel'dovich effects (tSZ/kSZ), the CIB, and radio galaxies. Additionally, low-redshift maps of cosmic shear following the observational properties Dark Energy Survey Year 1 and Vera Rubin Observatory Year 1 data are also made public.

Since both \synsky and the fiducial \SkyLine trace the same structure and have their star formation following the same underlying model, the combination of these two suites of simulations is a tantalizing playground to study cross correlations between LIM surveys and other probes. To highlight some of the possibilities, we showcase some aspects of the cross-correlation between the 21 cm and Ly$\alpha$ surveys introduced in the previous subsection (now covering the full sky) with maps of the Compton-$y$ distortions produced by the tSZ effect at 176 GHz and the CIB at 545 GHz, respectively.

The tSZ effect is sourced by CMB photons scattering off hot gas in massive halos which primarily inhabit the low redshift $z\leq 1$ Universe. A combined analysis of the tSZ and different emission lines can then probe different phases of the intergalactic medium, complementing LIM-only searches as explored in \citet{Sun:2020mco}. On the other hand, at 545 GHz, most of the contribution to the CIB comes from emission of UV-heated dust between redshifts $1 \leq z \leq 5$, peaking at $z\approx 2-3$ (see e.g., \citet{Planck:2015emq}), which corresponds to a similar redshift range probed by LIM experiments. Therefore, combining LIM and CIB observations will complement studies of the star formation history in cross correlations of the CIB with galaxies \citep{Jego:2022eqo, Jego:2022fkl} extending them to higher redshifts. Combining LIM and CIB entails further benefits, including breaking the degeneracies between the parameters modeling the SFR and those connecting it with the line luminosity \citep{Zhou:2022gmu}, and providing a statistical reconstruction of the whole SED of galaxies and intergalactic medium in the optical and infrared, combining continuum and line emissions. 

\begin{figure}
    \centering
    \includegraphics[width=\linewidth]{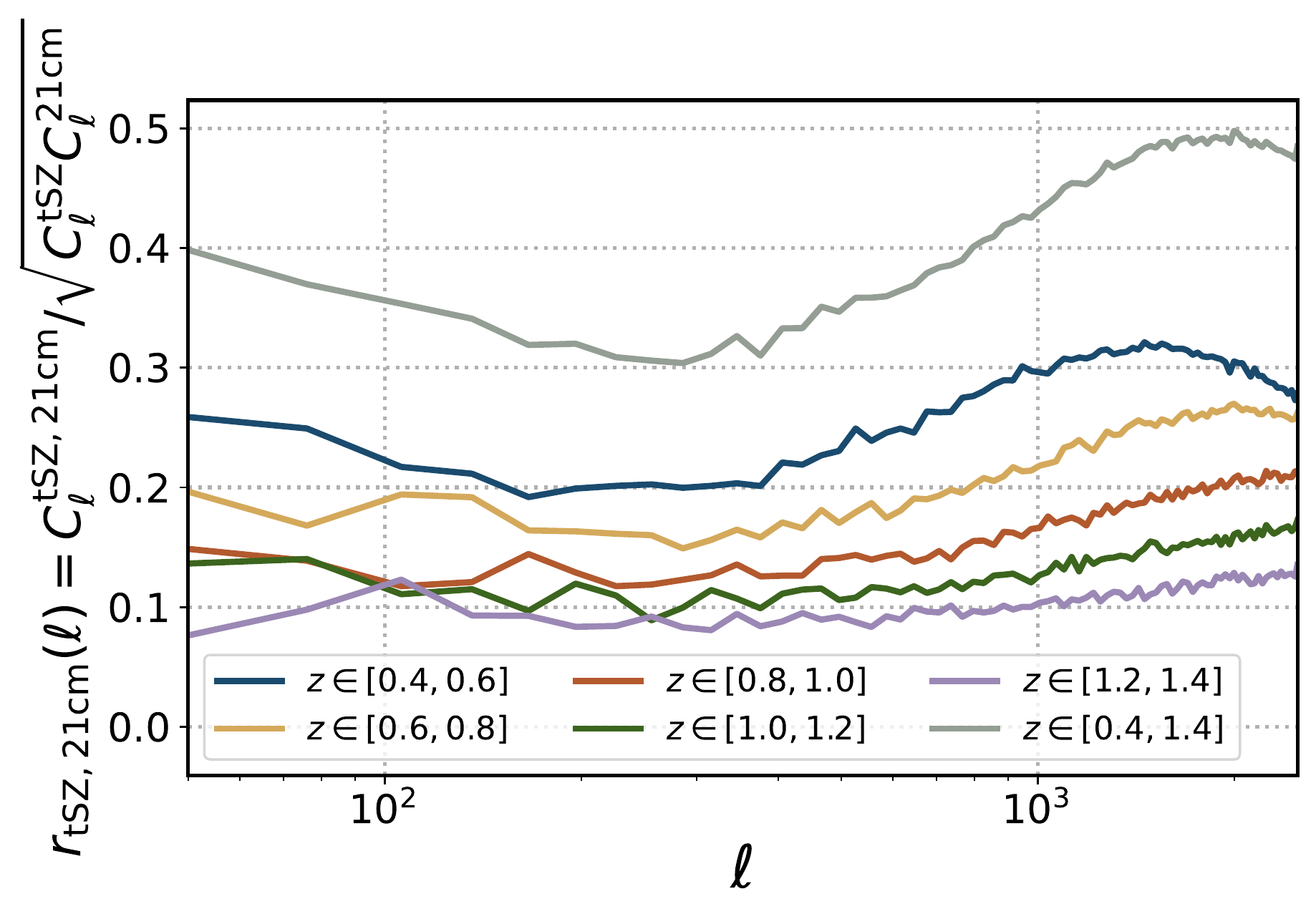}
    \includegraphics[width=\linewidth]{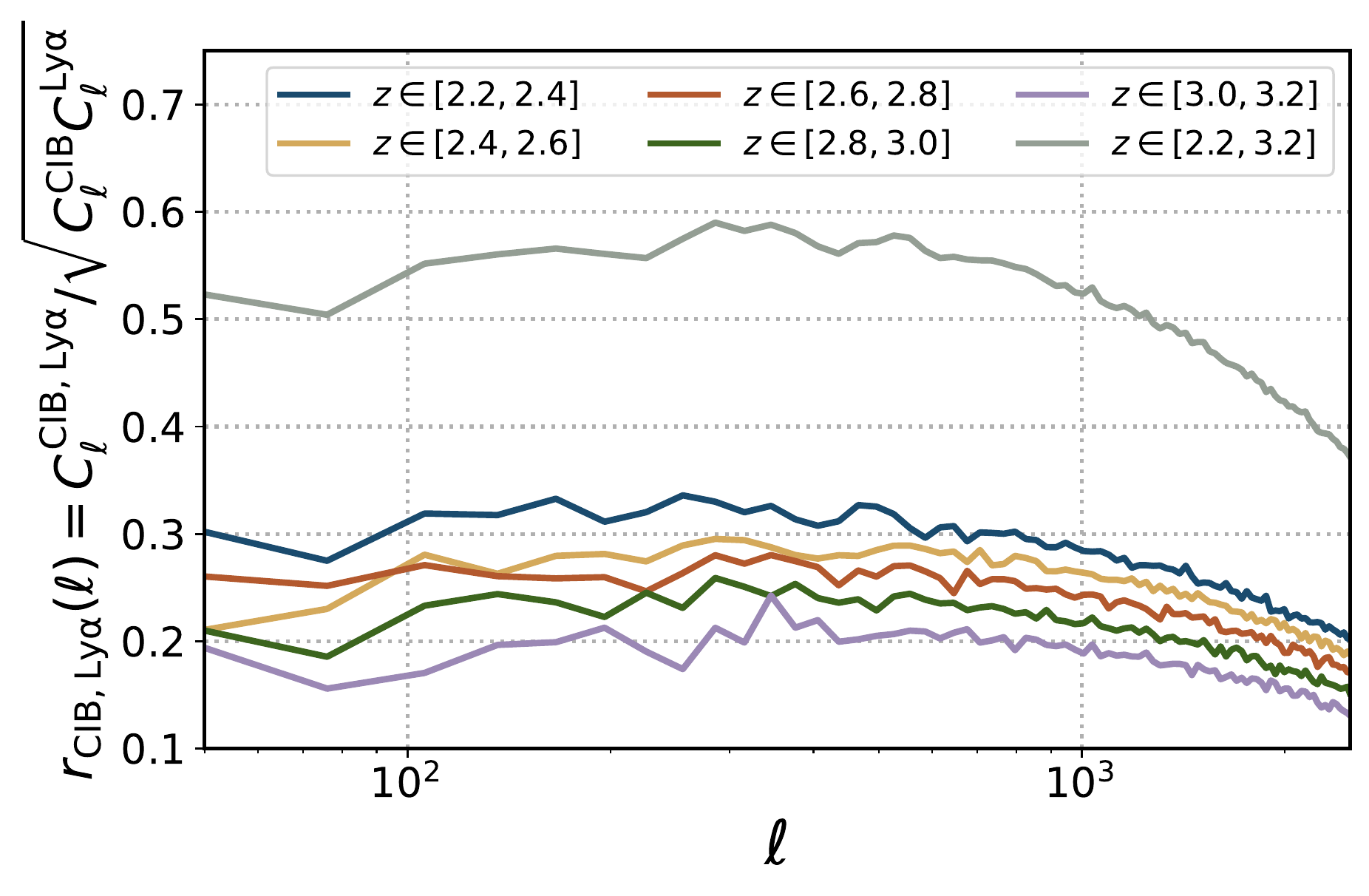}
    \caption{Angular cross-correlation coefficient between 21cm intensity maps and tSZ at 176 GHz (top) and Ly$\alpha$ intensity mapping and CIB at 545 GHz (bottom) for full sky maps. Different colors correspond to different redshift binning of the line-intensity maps}
    \label{fig:cross_cmbsec}
\end{figure}

We show the angular cross-correlation coefficients for both cases in Fig.~\ref{fig:cross_cmbsec}. Although these simulated maps are full sky, MDPL2 is a 1 (Gpc/$h)^3$ box. Therefore, the largest angular modes fall outside the size of the simulation used to build the lightcone. To mitigate this we only consider $\ell>50$ in Fig.~\ref{fig:cross_cmbsec}. We bin the power spectra in bins of $\Delta\ell=30$, taking the mean $C_\ell$ and the mid point in $\Delta\ell$ as the data to plot. 

Observations of the tSZ and the CIB return integrated maps which can be tomographically explored in cross correlation with an observable with line-of-sight information, as is the case of LIM. Therefore, we bin in redshift the LIM maps introduced in \S~\ref{sec:limxgals} in $\Delta z=0.2$ width bins. As expected, the complete maps (with $\Delta z=1$) is more correlated than the binned maps, and in no case the cross correlation reaches unity because the tSZ and CIB maps contain information from redshifts outside the range contained in the line-intensity maps. The tSZ contribution peaks at low redshift and decreases at $z\gtrsim 0.4$. This is why we see how the cross correlation with the 21 cm maps decreases with redshift. Also, the correlation grows at small scales because the tSZ signal is dominated by the 1-halo term. As mentioned before, the CIB intensity follows the star-formation rate density. This is why we find a decreasing cross-correlation coefficient as we increase the redshift of the Ly$\alpha$ maps. Finally, the cross correlation between the CIB and the Ly$\alpha$ maps decreases at larger scales than the tSZ because the angular smoothing corresponds to larger physical scales at higher redshifts and the CIB does not present such strong contributions from the 1-halo term. 

{The examples shown in this discussion do not include all the complications introduced by the presence of continuum foregrounds in the observations. This contribution to the map is usually much brighter than the desired signal, and smooth in frequency (see \S~\ref{sec:foregrounds}). The lack of precise enough models for these foregrounds motivate the use of blind foreground removal methods, which effectively remove the longest modes parallel to the line of sight (see e.g.,~\citet{Switzer:2015ria, Switzer:2018tel, Cunnington:2023jpq}). Since most of the information in the cross-correlation of projected fields like secondary CMB anisotropies and the CIB lie in the longest parallel modes, foreground removal may severely impact the results shown in Fig.~\ref{fig:cross_cmbsec}. The unique characteristics of \SkyLine and the coherence with the simulated maps from \synsky makes the combination of these tools a promising sandbox to develop strategies to overcome such challenges.} 

\section{Conclusions}
\label{sec:conclusion}

LIM is poised to deliver unprecedented measurements of large-scale structure at high redshifts, while providing invaluable information about galaxy evolution through its sensitivity to faint and diffuse sources and by mapping the different phases of the interstellar and intergalactic media. However, observational contamination challenges the detection of a cosmological signal, and the high level of non-Gaussianity of the maps limits the interpretation of the measurements. Therefore, it is of key importance to develop realistic mocks to prepare the tools to analyze forthcoming LIM measurements, characterize the physical information within line-intensity maps, and maximize the return from LIM surveys. 

This work presents a framework to quickly generate LIM lightcones for any spectral line (except for HI during reionization), as well as line interlopers, galactic foregrounds, and instrument noise, with a potential extension to include extragalactic continuum emission left for future work. By starting from a galaxy catalog, this scheme also enables us to build self-consistent maps of quenched and star-forming galaxy populations, which can be used to explore the cross-correlation between tracers. An implementation of this framework is presented in a public code, \SkyLine. The fiducial implementation in this work possesses the same large-scale structure as the \synsky simulated skies from~\citet{Omori:SimSky}, providing an accurate and realistic simulated sky to cross-correlate LIM measurements and other observables, including galaxy clustering, galaxy weak lensing, CIB and CMB secondary anisotropies. 

The approach presented in this work is unique in terms of flexibility, accuracy in the halo distribution and redshift evolution of the clustering and signal, opportunities of cross correlations with other observables, and inclusion of observational contaminants. Moreover, its modular structure allows for a straightforward addition of other spectral lines or emission models, as well as alternative initial halo distributions and astrophysical properties. 

In this work we have described the structure of \SkyLine and showcased with some examples its ample set of possibilities. These include the study of population of emitters that are not directly observable with LIM but that can be used to calibrate astrophysical line-emission models and improve the characterization of the information content in line-intensity maps, measurements of the anisotropic power spectrum and the VID, the inclusion of line interlopers and galactic foregrounds, and cross correlations with galaxies, the CIB and the tSZ. 

{This work is complementary to recent efforts to simulate LIM observations on a lightcone, such as those presented in~\citet{Bethermin:2022lmd},~\citet{Gkogkou:2022bzo},  or~\citet{2021ApJ...911..132Y}, which rely on semi-empirical or semi-analytic models to assign astrophysical properties to galaxies. The aforementioned mocks focus on more sophisticated models of line emission in the sub-millimeter range (i.e., CO and [CII] lines, mainly) 
over narrow lightcones covering small patches on the sky. On the other hand, \SkyLine is highly flexible and is able to capture a larger number of lines in any frequency range simultaneously over a large volume and deep redshift range. Furthermore these signals are coherently simulated with maps of other tracers of large-scale structure. This combination of large volumes, multiple tracers, and the inclusion of a vast number of realistic observational effects fills a unique position in the space of LIM mocks. The fiducial catalogs and code are also public, making \SkyLine an ideal community tool.} 

The scheme to generate flexible, multi-tracer and multi-line LIM mocks we have introduced in this work can enable a wide variety of follow-up analyses that will enhance our understanding of LIM as a probe of large-scale structure and astrophysics in the coming decade. Mocks generated with our approach can be used to validate analysis codes for summary statistics of LIM surveys and their cross-correlations (see e.g., \cite{DES:2019jmj,DeRose_2022} for galaxy surveys). Our catalogs can also aid in the interpretation of LIM measurements, both through a breakdown of the properties of emitters that contribute to the signal (such as was done for the VID in  Fig.~\ref{fig:voxel_LF}) and by quantifying the influence that changes in astrophysical modelling have on the signal. Finally, we anticipate this will be a valuable tool to realistically explore the impact of observational contaminants and limitations on more futuristic measurements of large-scale structure performed with LIM.

\section*{acknowledgments}
We are grateful to Yuuki Omori for essential discussions, for coordinating the lightcone rotation matrices and underlying catalog, and  for sharing the maps from \synsky employed in \S~\ref{sec:cross_CMB}. We also thank Marc Kamionkowski, Ely Kovetz, Maja Lujan Niemeyer, and Risa Wechsler for useful discussions, and Dongwoo Chung for comments on the draft. GSP was supported by the National Science Foundation Graduate Research Fellowship under Grant No. DGE1746891.  N.K. acknowledges support from the Gerald J. Lieberman Fellowship for part of the duration of this work. JLB was supported by the Allan C. and Dorothy H. Davis Fellowship during part of the development of this project.  Figures and code to produce them in this work have been made using \texttt{nbodykit} \citep{Hand:2017pqn} and the SciPy Stack \citep{2020NumPy-Array,2020SciPy-NMeth,4160265}. This research has made use of NASA's Astrophysics Data System and the arXiv preprint server.
\section*{Data Availability}
The underlying data products used to create the fiducial realization of \SkyLine, \textsc{MDPL2-UM}, are publicly available at \href{https://www.peterbehroozi.com/data.html}{https://www.peterbehroozi.com/data.html}. The \SkyLine code and associated products will be made publicly available upon acceptance and are also available on reasonable request to the authors before acceptance.

\bibliography{limbib}
\bibliographystyle{mnras}
\appendix
\section{Impact of simulation resolution at high redshifts}
\label{appendix:A}
\begin{figure*}
    \centering
    \includegraphics[width=\linewidth]{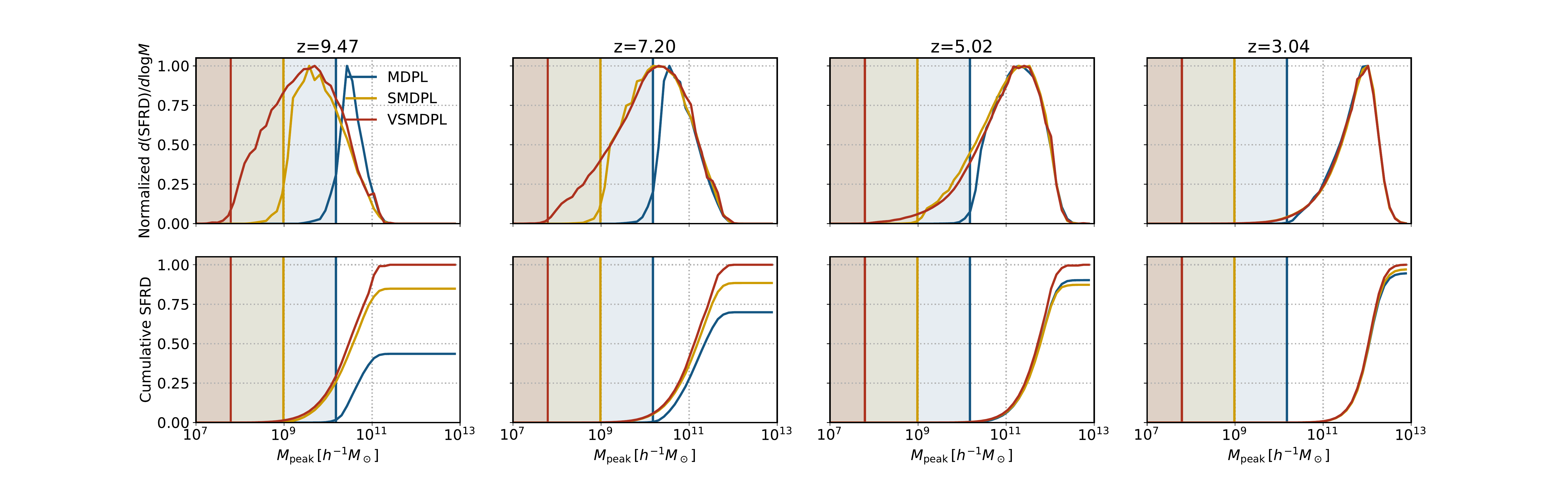}
    \caption{\emph{Top panel:} The star formation rate density as a function of peak halo mass. At higher masses all of the three suites of simulations agree, while for lower masses we can see the appreciable contribution to SFRD coming from unresolved halos in our fiducial simulation. The shaded bands correspond to the regime where halos have fewer than 10 particles, e.g. $M_{\rm peak} < 10 M_{\rm particle}$, for each simulation.  \emph{Bottom panel:} Cumulative star formation density (SFRD) for each simulation, as a function of peak halo mass, normalized to the final value from \textsc{VSMDPL-UM}, the highest resolution simulation. The difference between the different curves directly translates to the amount of SFRD that is lost, as a function of particle resolution, across redshift. At $z\sim9.5$ our fiducial simulation contains around 45\% of the total SFRD relative to our highest resolution box.}
    \label{fig:sfrd_resolution}
\end{figure*}
As discussed in \S~\ref{sec:halo_astro}, the limited mass resolution of our fiducial \textsc{MDPL2-UM} simulations can have an impact in our ability to properly resolve star formation (and therefore line emission) at higher redshifts. The causes are two-fold: halos that are resolved have unreliable SFR due to lower particle count, and the contribution from halos of masses below the resolution threshold to the SFR density could be non negligible. In this appendix we investigate this by assessing two catalogs related to \textsc{MDPL2-UM}: \textsc{SMDPL-UM} and \textsc{VSMDPL-UM}\footnote{Which stand for \textsc{Small} and \textsc{Very Small} \textsc{MultiDark Planck}, respectively. Information about these simulations can be found at \href{https://www.cosmosim.org/}{https://www.cosmosim.org/}. The \universemachine value-added catalogs are available at \href{https://www.peterbehroozi.com/data.html}{https://www.peterbehroozi.com/data.html}.}. These are comparable \universemachine+$N$-body catalogs, with the same cosmology and $N=3840^3$ particles, but at smaller volumes of $V=(400\, h^{-1} {\rm Mpc})^3$ and $(160\, h^{-1} {\rm Mpc})^3$, respectively. \par 
For each simulation we select the snapshots closest to $z\sim[9.5, 7.2, 5, 3]$ to compare the SFRD as function of peak halo mass and the cumulative SFRD from each simulation as function of redshift. We show the results in Fig.~\ref{fig:sfrd_resolution}: at $z\sim 9.5$ the fiducial suite can only track $\sim 45\%$ of the total SFR in the Universe, however by $z\sim3$ the simulations are entirely comparable; at $z\sim 5$, our fiducial simulation lacks $\sim 10\%$ of the total SFR density. This implies, then, that our high-redshift line luminosities at $z \gtrsim 5$ are not as robustly modelled as they could be. 

However, we caution that the results on SFR density do not directly translate to line luminosity. Each line luminosity has a different relation with the SFR, and at lower halo masses its emission can be suppressed due to additional astrophysical complications, including the possibility that galaxies populating these halos may be less chemically evolved or less dusty. This limitation is common to all simulation-based analyses and generation of mocks maps (see e.g.,~\citet{Bethermin:2022lmd}). 
A complete assessment of the impact of simulation resolution on line luminosity as a function of redshift significantly depends on the line and models of interest, and it is then beyond the scope of this work. For studies related to specific surveys and spectral lines, a tailored analysis should be performed, and other simulations than our fiducial, such as \textsc{SMDPL-UM} or \textsc{VSMDPL-UM} may be required. 


\bsp	
\label{lastpage}
\end{document}